\documentclass[%
 reprint,
 superscriptaddress,
 amsmath,amssymb,
 aps,
 prfluids, 
 onecolumn,
 floatfix
]{revtex4-2}

\usepackage{graphicx}
\usepackage{siunitx}

\usepackage{tikz}

\usepackage[usenames,dvipsnames]{xcolor}
\newcommand{\nll}{\not\ll}

\newcommand{\hypgeo}[2]{%
  \operatorname{%
    {\vphantom{\mathnormal{F}}}_{#1}%
    \kern-\scriptspace
    \mathnormal{F}_{#2}%
  }%
}

\usepackage[%
    linkcolor = blue, 
    citecolor = red, 
    urlcolor = blue, 
    colorlinks = true
]{hyperref}

\begin{document}

\title{Viscoelastic flow of an Oldroyd-B fluid through a slowly varying contraction-expansion channel: pressure drop and elastic stress relaxation}

\author{Yali Kedem}
\affiliation{Faculty of Mechanical Engineering, Technion -- Israel Institute of Technology, Haifa 3200003, Israel}
\author{Bimalendu Mahapatra}
\affiliation{Faculty of Mechanical Engineering, Technion -- Israel Institute of Technology, Haifa 3200003, Israel}
\affiliation{Department of Mechanical Engineering, Indian Institute of Science Bangalore, Karnataka 560012, India}
\author{Evgeniy Boyko}
\email[Email address for correspondence: ]{evgboyko@technion.ac.il}
 \affiliation{Faculty of Mechanical Engineering, Technion -- Israel Institute of Technology, Haifa 3200003, Israel}

\date{\today}

\begin{abstract}
Flows of viscoelastic fluids in narrow non-uniform geometries, such as contraction, expansion, and contraction-expansion configurations, are ubiquitous in various engineering applications and physiological flow systems. For such flows, one of the key interests lies in understanding how fluid viscoelasticity affects the dependence of the pressure drop on the flow rate, which remains not fully understood compared to Newtonian flows.
In this work, we analyze the flow of the Oldroyd-B fluid in slowly varying contraction-expansion channels, commonly referred to as constrictions.
Unlike most previous theoretical studies that focus on contracting channels, we consider instead a constriction geometry and present a theoretical model for calculating the elastic stresses and flow rate-pressure drop relation at low and high Deborah ($De$) numbers.
We apply lubrication
theory in orthogonal curvilinear coordinates and consider the ultra-dilute limit, in which the velocity approximates a parabolic and Newtonian profile. This results in a one-way coupling between the velocity and elastic stresses, allowing us to derive closed-form expressions for the elastic stresses and pressure drop for arbitrary values of the Deborah number.
We validate our theoretical predictions with two-dimensional numerical simulations and find excellent agreement.
We identify the physical mechanisms governing the pressure drop behavior and compare our results for the slowly varying constriction with previous predictions for the contraction. We reveal that, at low $De$, the pressure drop in the constriction monotonically decreases with $De$, similar to the contraction geometry. 
However, at high
Deborah numbers, in contrast to a linear decrease for the contraction, the pressure drop across the constriction reaches a plateau due to the vanishing contribution of elastic normal stresses, leaving elastic shear stresses as the sole driver of the reduction. Furthermore, we elucidate the spatial relaxation of elastic stresses and pressure gradient in the exit channel following both constriction and contraction geometries, showing that the relaxation length is significantly shorter in the case of a constriction.
 \end{abstract}


\maketitle

\section{Introduction}

In recent years, the pressure-driven flow of viscoelastic fluids through narrow, slowly varying geometries has attracted considerable attention in the fluid mechanics and rheology communities due to its importance in industrial processes~\cite{pearson,tadmor2013principles}, microfluidic applications~\cite{ober2013microfluidic}, and physiological flows~\cite{westein2013atherosclerotic}. Viscoelastic fluid flows can exhibit markedly different behavior compared to Newtonian flows, even in the presence of a low concentration of polymer molecules~\citep{bird1987dynamics1,steinberg2021elastic,datta2022perspectives,ewoldt2022designing}.
Polymer molecules in the solution give rise to complex rheological phenomena, such as normal stress differences and extensional thickening, which significantly influence the hydrodynamic features of the flow, including the relationship between the pressure drop $\Delta p$ and the flow rate $q$.
In contrast to viscous Newtonian fluids flowing through non-uniform geometries, where the pressure drop exhibits a linear dependence on the flow rate at low Reynolds numbers, viscoelastic fluids display significantly different behavior, characterized by a nonlinear relationship between pressure drop and flow rate~(see, e.g.,~\citep{szabo1997start,rothstein1999extensional,rothstein2001axisymmetric,alves2003benchmark,zografos2020viscoelastic,boyko2021RT,boyko2022pressure,varchanis2022reduced,housiadas2023lubrication,boyko2024flow,hinch2024fast,Housiadas_Beris_2024,mahapatra2025viscoelastic}).

The relationship between the pressure drop and the flow rate in viscoelastic fluids has been studied extensively through both numerical simulations~\cite{szabo1997start,alves2003benchmark,binding2006contraction,alves2007divergent,zografos2020viscoelastic,alves2021numerical,varchanis2022reduced,zografos2022viscoelastic} and experimental measurements~\cite{rothstein1999extensional,rothstein2001axisymmetric,sousa2009three,campo2011flow,ober2013microfluidic,james2021pressure} in various rapidly varying geometries, including abrupt or hyperbolic contraction and contraction-expansion (constriction) channels.
We refer the reader to the recent overviews provided by~\citet{boyko2022pressure} and~\citet{hinch2024fast}. Such rapidly varying geometries often feature sharp corners, which greatly complicate theoretical analysis. Therefore, to enable asymptotic analysis, recent theoretical studies have considered instead slowly varying geometries, allowing for the application of lubrication theory~\cite{boyko2022pressure,housiadas2023lubrication,boyko2024flow,hinch2024fast,Housiadas_Beris_2024,housiadas2024pressure,mahapatra2025viscoelastic,sialmas2025exact,sialmas2025general}.

To date, most theoretical studies on the pressure-driven flow of viscoelastic fluids have focused on slowly varying contracting channels using lubrication theory~\cite{boyko2022pressure,housiadas2023lubrication,boyko2024flow,hinch2024fast,Housiadas_Beris_2024,housiadas2024pressure,mahapatra2025viscoelastic,sialmas2025exact,sialmas2025general}. To obtain analytical results, many studies considered the weakly viscoelastic limit by applying a
perturbation expansion in powers of the Deborah number $De$ (see definition in Sec.~\ref{Scaling})~\cite{boyko2022pressure,housiadas2023lubrication,Housiadas_Beris_2024,mahapatra2025viscoelastic,sialmas2025general}. For example, \citet{boyko2022pressure} studied the steady flow of an Oldroyd-B fluid in a slowly varying planar contraction and provided asymptotic expressions for the dimensionless pressure drop up to $O(De^3)$ in the low-$De$ limit.~\citet{housiadas2023lubrication} extended this low-$De$ analysis to much higher asymptotic orders and provided asymptotic expressions for the pressure drop up to $O(De^8)$ for different constitutive models, including Oldroyd-B, Giesekus~\citep{giesekus1982simple}, Phan-Thien$-$Tanner (PTT)~\citep{thien1977new,phan1978nonlinear}, and a finitely extensible nonlinear elastic (FENE) model with the Peterlin approximation (FENE-P)~\citep{bird1980polymer,bird1987dynamics1}. 
Nevertheless, their low-$De$ theoretical predictions for pressure drop, based on more complex constitutive equations, were in close agreement with those of the Oldroyd-B model and showed a monotonic decrease in the pressure drop with $De$ for the flow through a hyperbolic contraction.  
Recently,~\citet{sialmas2025general} demonstrated excellent agreement between the analytical predictions, based on high-order, low-Deborah-number asymptotic expressions of~\citet{housiadas2023lubrication}, and the finite-volume numerical simulations of \citet{mahapatra2025viscoelastic} for the pressure drop of an Oldroyd-B fluid flowing through a contracting hyperbolic channel for order-one Deborah numbers.
Furthermore,~\citet{Housiadas_Beris_2024} studied the flow of an Oldroyd-B fluid through an axisymmetric hyperbolic pipe in the low-$De$ limit. They derived asymptotic expressions for the pressure drop up to $O(De^8)$, thereby extending the earlier analyses by~\citet{housiadas2023lubrication} and~\citet{housiadas2024pressure}. Complementing this low-$De$ approach,~\citet{sialmas2025exact} developed an analytical solution for the pressure drop that remains valid at moderate Deborah numbers.

Beyond pressure-driven flows of viscoelastic fluids in slowly varying geometries, lubrication theory combined with a perturbation expansion in powers of the Deborah number has been applied extensively to the study of various
viscoelastic fluid flows. This includes viscoelastic thin-film boundary-driven lubrication~\cite{tichy1996non,sawyer1998non,ahmed2021new,gamaniel2021effect,ahmed2023modeling,sari2024effect, sari2025role,AhmedBiancofiore2025}, free-surface flows~\cite{ro1995viscoelastic,zhang2002surfactant,saprykin2007free,datt2022thin}, and motion of a sphere near a rigid plane in a viscoelastic fluid~\cite{Ardekani_2007,Ruangkriengsin2024}. For example,~\citet{AhmedBiancofiore2025} recently studied the interplay between the fluid viscoelasticity and the sidewalls of finite-width channels in thin-film lubrication contacts using the Oldroyd-B model in the low-Deborah-number limit.

While the low-Deborah-number asymptotic analysis often provides closed-form analytical expressions, its validity remains limited, and it cannot accurately capture the behavior of elastic stresses and pressure drop at high Deborah numbers.~One theoretical approach to studying viscoelastic flows at non-small Deborah numbers is to consider the ultra-dilute limit, which physically corresponds to a low concentration of polymer molecules in a Newtonian solvent~\citep{remmelgas1999computational,koch2016stress,moore2012weak,li2019orientation,mokhtari2022birefringent,sharma2025extensional,boyko2024flow,hinch2024fast,mahapatra2025viscoelastic}. In the ultra-dilute limit, it is assumed that the contribution of the polymer viscosity $\mu_p$ to the total zero-shear-rate viscosity of the polymer solution $\mu_0$ is small, specifically that $\beta_p = \mu_p/\mu_0 \ll 1$. As a result, in the ultra-dilute limit, the flow field can be approximated as Newtonian, leading to the generation of elastic stresses that are not coupled back to change the flow at the leading order in $\beta_p$~\citep{remmelgas1999computational,moore2012weak,li2019orientation,mokhtari2022birefringent,sharma2025extensional}. Such a one-way coupling greatly simplifies theoretical analysis, allowing for the use of elastic stresses to determine corrections to the velocity and pressure fields due to fluid viscoelasticity, even at high Deborah numbers.

\citet{boyko2024flow} and \citet{hinch2024fast} have recently analyzed the flow of an Oldroyd-B fluid through a slowly varying planar contraction at high Deborah numbers using lubrication theory. \citet{boyko2024flow}~considered the ultra-dilute limit and derived semi-analytical expressions for the conformation tensor and pressure drop for arbitrary values of the Deborah number. Furthermore,~\citet{boyko2024flow} and~\citet{hinch2024fast} provided asymptotic solutions in the high-$De$ limit, demonstrating that, for the contraction geometry, the pressure drop of the Oldroyd-B fluid monotonically decreases
with increasing $De$, and at high Deborah numbers, the pressure drop scales linearly with $De$. They also identified two distinct physical mechanisms, based on elastic normal and shear stresses, that are responsible for this pressure drop reduction. 
In addition to the pressure drop in the contraction,~\citet{boyko2024flow} studied the relaxation of elastic stresses and pressure gradient in the exit channel following the contraction, showing that the downstream distance required for such relaxation scales linearly with $De$. Understanding this relaxation length is of both fundamental and practical importance, as it sets the appropriate size of the computational domain~\cite{alves2003benchmark}.
Recently,~\citet{mahapatra2025viscoelastic} extended the 
analysis of~\citet{boyko2024flow} and \citet{hinch2024fast} to the FENE-CR model introduced by~\citet{chilcott1988creeping}, elucidating how the finite extensibility of polymer chains influences the elastic stresses and pressure drop in a slowly varying planar contraction.~\citet{mahapatra2025viscoelastic}  showed that at high $De$, unlike a linear decrease for the Oldroyd-B fluid, the pressure drop of the FENE-CR fluid exhibits a non-monotonic variation due to finite extensibility, first decreasing and then increasing with $De$.

In contrast to the extensively studied slowly varying contracting channels, the slowly varying contraction-expansion geometry, also referred to as a constriction, has received less attention in the fluid mechanics community. Recently, \citet{hinch2024fast}~studied numerically the flow of the Oldroyd-B fluid through a slowly varying constriction for order-one Deborah numbers, and derived asymptotic solutions at high
Deborah numbers based on the ultra-dilute limit, $\beta_p\ll1$.
They solved numerically the viscoelastic lubrication equations and presented results for $\beta_p = 0.5$ at moderate Deborah numbers, demonstrating that the pressure drop across the constriction decreases with increasing $De$.
However, in contrast to the pressure drop plateau predicted by the high-$De$ asymptotic analysis, their numerical results showed no sign of leveling off even at the highest Deborah numbers tested. In the current study, we rationalize this discrepancy and show theoretically that, in the ultra-dilute limit, the pressure drop of the Oldroyd-B fluid across the constriction levels off to a plateau at high Deborah numbers, in excellent agreement with the high-$De$ asymptotic prediction.

In this work, we study the pressure-driven flow of the Oldroyd-B fluid in slowly varying planar constriction channels at low and high Deborah numbers. 
In contrast to~\citet{hinch2024fast}, who studied numerically the flow through a constriction for order-one Deborah numbers, in the current work, we consider the ultra-dilute limit, which enables us to explore arbitrary values of Deborah number. Specifically, we analyze the pressure drop of an Oldroyd-B fluid in a constriction geometry, as well as the relaxation of elastic stresses and the pressure gradient in the exit channel, thereby extending the semi-analytical framework of~\citet{boyko2024flow}, originally developed for contracting channels.
Furthermore, we elucidate the physical mechanisms governing the pressure drop behavior and compare our results for the slowly varying constriction with previous predictions for the contraction.
Due to the well-known numerical challenges at high Deborah numbers, commonly referred to as the high-Weissenberg-number problem~(see, e.g.,~\citep{owens2002computational, alves2021numerical}), we believe that our analytical and semi-analytical results in the ultra-dilute limit, which are valid at high $De$, are essential for assessing the accuracy of
simulation predictions, comparing with experimental measurements, and deepening our understanding of viscoelastic channel flows.

The paper is organized as follows. In Sec.~\ref{PF}, we present the problem formulation and the dimensional governing equations for the pressure-driven flow
of the Oldroyd-B fluid. We further identify the key dimensionless
parameters governing the flow and provide the non-dimensional lubrication equations and boundary conditions in orthogonal curvilinear coordinates.
In Sec.~\ref{Ultra-dilute section}, we present lubrication analysis in the ultra-dilute limit and provide semi-analytical expressions for the conformation
tensor in the constriction for arbitrary values of the Deborah number. These semi-analytical expressions allow us to calculate the pressure drop and elucidate the elastic normal and shear stress contributions to the pressure drop for all $De$. 
In Sec.~\ref{Results}, we present our theoretical predictions for the elastic stresses, pressure drop, and the respective contributions of elastic stresses to the pressure drop within the constriction, demonstrating excellent agreement with the high-$De$ asymptotic results.
We validate the semi-analytical results for the pressure drop by comparing them with two-dimensional finite-volume numerical simulations, finding excellent agreement.
We further compare and contrast the predictions for constriction and contraction, highlighting the differences in the physical mechanisms that govern the pressure drop behavior and the relaxation in the exit channel.
Finally, we conclude with a discussion of the results in Sec.~\ref{Concluding_remarks}.

\section{Problem formulation and governing equations}\label{PF}

\begin{figure}
\centerline{\includegraphics[scale=1.15]{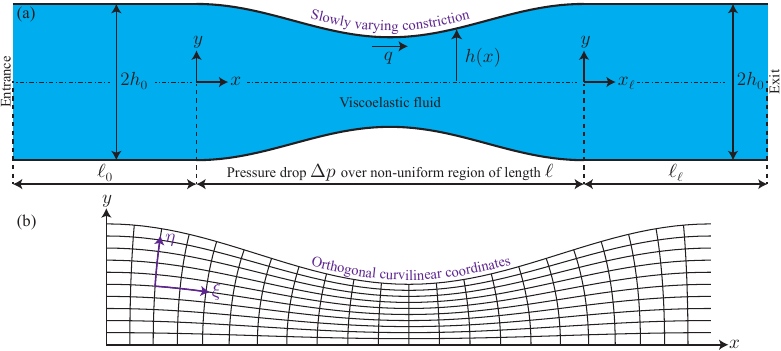}}
\caption{(a) Schematic illustration of the planar configuration consisting of a slowly varying and symmetric contraction-expansion channel of height $2h(x)$ and length $\ell$ $(h\ll \ell)$, connected to two long straight channels of height
$2h_0$ up- and downstream. The contraction-expansion channel, which we refer to as constriction, contains a viscoelastic fluid steadily driven by the imposed flow rate $q$. 
(b) Schematic illustration of the orthogonal curvilinear coordinates ($\xi,\eta$) used in the lubrication analysis of a slowly varying constriction. The coordinate $\xi$ is constant along vertical grid lines, and $\eta$, defined in (\ref{curvilinear mapping}), is constant along
the curves going from left to right. We are interested in determining the pressure drop $\Delta p$ over the constriction region and the spatial relaxation of
pressure and elastic stresses in the exit channel.}\label{F1}
\end{figure}

We study the incompressible steady flow of a 
viscoelastic fluid in a slowly varying
and symmetric two-dimensional constriction of height $2h(x)$ and length $\ell$, where $h(x)\ll\ell$, as shown in Fig.~\ref{F1}(a). 
Simiar to previous experimental, numerical, and theoretical studies~(see, e.g.,~\citep{szabo1997start,rothstein1999extensional,alves2003benchmark,campo2011flow,ober2013microfluidic,zografos2020viscoelastic,boyko2022pressure,housiadas2023lubrication,boyko2024flow,hinch2024fast,Housiadas_Beris_2024,mahapatra2025viscoelastic}), we assume that the inlet ($x=0$) and outlet ($x=\ell$) of the constriction are connected to two long straight channels of height $2h_{0}$, and length $\ell_{0}$ and $\ell_{\ell}$, respectively.
We consider the fluid motion with the velocity $\textbf{u}$, stress tensor $\boldsymbol{\sigma}$, and pressure distribution $p$ induced by an imposed flow rate $q$ (per unit depth). Our primary interest in this work is to examine the pressure drop $\Delta p$ of a viscoelastic fluid over the constriction region and the spatial relaxation of pressure and elastic stresses in the exit channel at low and high Deborah numbers.

For our analysis, it is convenient to use two different coordinate systems. The first is Cartesian coordinates $(x,y)$ and $(x_{\ell},y)$, where the $x$ and $x_{\ell}=x-\ell$ axes lie
along the symmetry midplane of the channel (dashed-dotted line) and $y$ is in the direction of the shortest dimension. We denote the velocity components in Cartesian coordinates as $\textbf{u} =u_x\textbf{e}_x+u_y\textbf{e}_y$.
The second coordinate system is orthogonal curvilinear coordinates $(\xi,\eta)$ defined in (\ref{curvilinear mapping}) and illustrated in Fig.~\ref{F1}(b). We denote the velocity components in orthogonal curvilinear coordinates as $\textbf{u}=u\textbf{e}_\xi+v\textbf{e}_\eta$.

We consider low-Reynolds-number flows so that the fluid motion is governed by the continuity equation and Cauchy momentum equations in the absence of inertia
\refstepcounter{equation}
$$
\boldsymbol{\nabla}\cdot \textbf{u}=0,\qquad\boldsymbol{\nabla}\cdot\boldsymbol{\sigma}=\boldsymbol{0}.\eqno{(\theequation{\text{a,b}})}\label{Continuity+Momentum}
$$

We describe the viscoelastic behavior of the fluid using the Oldroyd-B constitutive model~\citep{oldroyd1950formulation}. 
This model represents the simplest combination of viscous and elastic stresses, making it widely used to describe the flow of viscoelastic Boger fluids with a constant shear viscosity~\citep{james2009boger}. The Oldroyd-B equation can be derived from microscopic principles by modeling polymer molecules as elastic dumbbells, which follow a linear Hooke's spring law for the restoring force as they are advected and stretched by the flow~\citep{bird1987dynamics1,larson1988constitutive}. 
The stretching of the polymer molecules beyond their equilibrium state results in polymer stresses $\boldsymbol{\tau}_{p}$ due to the tension in the spring. 

For the Oldroyd-B fluid model, the stress tensor $\boldsymbol{\sigma}$ is given by
\begin{equation}
\boldsymbol{\sigma}=-p\textbf{I}+2\mu_{s}\textbf{E}+\boldsymbol{\tau}_{p}=-p\textbf{I}+2\mu_{s}\textbf{E}+\frac{\mu_{p}}{\lambda}(\textbf{A}-\textbf{I}).\label{Stress tensor}
\end{equation}
The first term on the right-hand side of~(\ref{Stress tensor}) is the pressure contribution, and the second term is the viscous stress contribution
of a Newtonian solvent with a constant viscosity $\mu_{s}$, where $\textbf{E}=(\boldsymbol{\nabla}\textbf{u}+(\boldsymbol{\nabla}\textbf{u})^{\mathrm{T}})/2$ is the rate-of-strain tensor. The last term is the polymer contribution to the stress tensor $\boldsymbol{\tau}_{p}$, which can be expressed in terms of the conformation tensor (or the deformation of the microstructure) $\textbf{A}$ as $\boldsymbol{\tau}_{p}=(\mu_{p}/\lambda)(\textbf{A}-\textbf{I})$~\citep{bird1987dynamics1,larson1988constitutive,Intro_C_F}, where $\lambda$ is the relaxation time and $\mu_{p}$ is the polymer contribution to the shear viscosity at zero shear rate. For convenience, we also introduce the total zero-shear-rate viscosity $\mu_0=\mu_s+\mu_p$.

The evolution equation for the conformation tensor $\textbf{A}$ of the Oldroyd-B model fluid
is given at steady state as~\citep{bird1987dynamics1,larson1988constitutive,Intro_C_F}
\begin{equation}
\textbf{u}\cdot\boldsymbol{\nabla}\textbf{A}-(\boldsymbol{\nabla}\textbf{u})^{\mathrm{T}}\cdot\textbf{A}-\textbf{A}\cdot(\boldsymbol{\nabla}\textbf{u})=-\frac{1}{\lambda}(\textbf{A}-\textbf{I}).\label{Evolution OB}
\end{equation}
Equations (\ref{Continuity+Momentum})--(\ref{Evolution OB}) form the basis of the governing equations for the pressure-driven flow of the Oldroyd-B fluid we consider in this work.

\subsection{Scaling analysis and non-dimensionalization}\label{Scaling}

We study the viscoelastic fluid flow through a narrow, slowly varying channel, in which $h(x) \ll \ell$. Therefore, to non-dimensionalize our fluid mechanical
problem, we introduce non-dimensional variables based on lubrication theory~\cite{tichy1996non,zhang2002surfactant,saprykin2007free,ahmed2021new,boyko2021RT,ahmed2023modeling,boyko2022pressure,housiadas2023lubrication,boyko2024flow,hinch2024fast,mahapatra2025viscoelastic,boyko2025pressureFD}
\begin{subequations}
\begin{gather}
(X,Y) = \left( \frac{x}{\ell}, \frac{y}{h_0} \right), \quad
(U_x,U_y) = \left( \frac{u_x}{u_c}, \frac{u_y}{\epsilon u_c} \right), \quad 
H(X) = \frac{h(x)}{h_0}, \quad 
P = \frac{p}{\mu_0 u_c \ell / h_0^2}, \quad 
\Delta P = \frac{\Delta p}{\mu_0 u_c \ell / h_0^2},\\
\mathcal{A}_{xx} = \epsilon^2 A_{xx}, \quad 
\mathcal{A}_{xy} = \epsilon A_{xy}, \quad 
\mathcal{A}_{yy} = A_{yy}, \\
\mathcal{T}_{p,xx} = \frac{h_{0}^2}{\mu_0 u_c \ell} \tau_{p,xx}, \quad
\mathcal{T}_{p,xy} = \frac{h_0}{\mu_0 u_c} \tau_{p,xy}, \quad 
\mathcal{T}_{p,yy} = \frac{\ell}{\mu_0 u_c} \tau_{p,yy}.
\end{gather}
\label{ND_variables_OB}\end{subequations}where $h_{0}$ is the half-height at $x=0$, $u_{c}=q/2h_0$ is the characteristic velocity in the streamwise direction, $q$ is the imposed flow rate per unit
depth, and $ \epsilon=h_{0}/\ell$ is the aspect ratio of the configuration, which is assumed to be small, $\epsilon\ll1$. 
In addition, we introduce the viscosity ratio $ \beta_p$ that determines the contribution of the polymer viscosity to the total viscosity,
\begin{equation}
    \beta_p=\frac{\mu_p}{\mu_s+\mu_p}=\frac{\mu_p}{\mu_0},
    \label{viscosity ratio}
\end{equation}
and the Deborah and Weissenberg numbers,
\begin{equation}
    De=\frac{\lambda u_c}{\ell} \quad \text{and} \quad Wi=\frac{\lambda u_c}{h_0}. \label{De and Wi}
\end{equation}The Deborah number, $De$, is the ratio of the relaxation time of the fluid, $\lambda$, to the residence time in the constriction region, $\ell/u_c$~\citep{tichy1996non,zhang2002surfactant,saprykin2007free,ahmed2021new,boyko2022pressure,ahmed2023modeling,housiadas2023lubrication,boyko2024flow,hinch2024fast,Housiadas_Beris_2024,mahapatra2025viscoelastic,boyko2025pressureFD,AhmedBiancofiore2025}. The Weissenberg number, $Wi$, is the product of the relaxation time of the fluid, $\lambda$, and the characteristic shear rate of the flow, $u_c/h_0$. 
For steady flows, the Deborah and Weissenberg numbers are often identical. However, in steady lubrication flows of viscoelastic fluids, as in our case, the Weissenberg number is related to the Deborah number through $De=\epsilon Wi$. This distinction arises from the presence of two different characteristic length scales in lubrication flows.
As a result, for lubrication flows in narrow geometries with $\epsilon \ll 1$, the Deborah number can remain small even when $Wi = O(1)$. 
We note that under the lubrication non-dimensionalization (\ref{ND_variables_OB}), the Deborah number $De$ appears naturally in the governing equations (see Appendix~\ref{AppA}), rather than the Weissenberg number~\citep {boyko2022pressure,housiadas2023lubrication,boyko2024flow,hinch2024fast,mahapatra2025viscoelastic,sari2025role}. Therefore, we will not use the Weissenberg number.

\subsection{Non-dimensional lubrication equations in orthogonal curvilinear coordinates}

For completeness, in Appendix~\ref{AppA}, we present the non-dimensional lubrication equations in Cartesian coordinates (see, e.g.,~\citep{boyko2022pressure, housiadas2023lubrication,boyko2024flow,hinch2024fast,mahapatra2025viscoelastic}). From (\ref{Momentum y ND Cart}), it follows that
$P=P(X)+O(\epsilon^2)$, i.e., the
pressure is constant across a cross-section but varies along the $x$-direction.

For the theoretical study of viscoelastic flow in slowly varying channels, it is convenient to transform the geometry of the constriction from the Cartesian coordinates $(X, Y)$ to curvilinear coordinates $(\xi, \eta)$, as shown in Fig.~\ref{F1}(b). To this end, we use the mapping
\citep{boyko2024flow,hinch2024fast,mahapatra2025viscoelastic}
\begin{equation}
    \xi=X-\frac{1}{2}\epsilon^2\frac{H'(X)}{H(X)}(H(X)^2-Y^2)+O(\epsilon^4)  \quad \text{and} \quad \eta=\frac{Y}{H(X)}.
    \label{curvilinear mapping}
\end{equation}
Similar to previous studies~\cite{boyko2024flow,hinch2024fast,mahapatra2025viscoelastic}, we use the following notation for the dimensionless components of velocity and conformation tensor in curvilinear coordinates $(\xi,\eta)$
\begin{equation}
\boldsymbol{U}=U\textbf{e}_\xi+V\textbf{e}_\eta \quad \text{and} \quad \boldsymbol{\mathcal{A}}= \mathcal{A}_{11}\textbf{e}_\xi\textbf{e}_\xi+\mathcal{A}_{12}(\textbf{e}_\xi\textbf{e}_\eta+\textbf{e}_\eta\textbf{e}_\xi)+\mathcal{A}_{12}\textbf{e}_\eta\textbf{e}_\eta,
\end{equation}
so that the velocity and conformation tensor components in the Cartesian and curvilinear coordinates are related through
\begin{subequations}
\begin{gather}
U_x = U - \epsilon^2 \eta H'(\xi) V, \quad 
U_y =  \eta H'(\xi) U+V, \\
\mathcal{A}_{xx} = \mathcal{A}_{11} + O(\epsilon^2), \\
\mathcal{A}_{xy} = \mathcal{A}_{12} + \eta H'(\xi) \mathcal{A}_{11} + O(\epsilon^2), \\
\mathcal{A}_{yy} = \mathcal{A}_{22} + 2 \eta H'(\xi) \mathcal{A}_{11} + \eta^2 H'(\xi)^2 \mathcal{A}_{12} + O(\epsilon^2).
\end{gather}\label{velocity_conformation_curvilinear}\end{subequations}Note that, since the difference between the $\xi$- and $X$-directions is only of order $O(\epsilon^2)$, for convenience, we keep using $X$ rather than $\xi$ in curvilinear coordinates~\cite{boyko2024flow,mahapatra2025viscoelastic}.

Using the mapping (\ref{curvilinear mapping}), the governing equations (\ref{Continuity+Momentum})--(\ref{Evolution OB}) take the following form in curvilinear coordinates~\citep{boyko2024flow,hinch2024fast,mahapatra2025viscoelastic}
\begin{subequations}\begin{align}
        \frac{\partial (H U)}{\partial X}&+ \frac{\partial V}{\partial \eta} = 0, \\
        \frac{dP}{dX} = (1 - \beta_p) \frac{1}{H^2} \frac{\partial^2 U}{\partial \eta^2} + \frac{\beta_p}{De}& \left( \frac{1}{H} \frac{\partial (H \mathcal{A}_{11})}{\partial X} + \frac{1}{H} \frac{\partial \mathcal{A}_{12}}{\partial \eta} \right),\\
     U \frac{\partial\mathcal{A}_{11}}{\partial X} + \frac{V}{H} \frac{\partial\mathcal{A}_{11}}{\partial \eta}
    - 2 \frac{\partial U}{\partial X} \mathcal{A}_{11}&
    - \frac{2}{H} \frac{\partial U}{\partial \eta} \mathcal{A}_{12}
    = -\frac{1}{De} \mathcal{A}_{11},\\
    U \frac{\partial \mathcal{A}_{12}}{\partial X} + \frac{V}{H} \frac{\partial \mathcal{A}_{12}}{\partial \eta}
    - H \frac{\partial}{\partial X} \left( \frac{V}{H} \right) \mathcal{A}_{11}&
    - \frac{1}{H} \frac{\partial U}{\partial \eta} \mathcal{A}_{22}
    = -\frac{1}{De} \mathcal{A}_{12}, \\
     U \frac{\partial\mathcal{A}_{22}}{\partial X} + \frac{V}{H} \frac{\partial\mathcal{A}_{22}}{\partial \eta}
    - 2H \frac{\partial}{\partial X} \left( \frac{V}{H} \right) \mathcal{A}_{12}&
    + 2 \frac{\partial U}{\partial X} \mathcal{A}_{22}
    = -\frac{1}{De} (\mathcal{A}_{22} - 1).
    \end{align}
\label{Conformation OB}\end{subequations}
The governing equations are supplemented by the following boundary conditions on the velocity,
\begin{equation}
U(X,1) =0, \quad  V(X,1) = 0, \quad \frac{\partial U}{\partial \eta}(X,0) = 0, \quad 
H(X) \int_0^1 U(X,\eta)\,\mathrm{d}\eta = 1, 
\label{BC OB}
\end{equation}
and the conformation tensor components,
\begin{equation}
\mathcal{A}_{11}(0,\eta)=\frac{18De^2}{H(0)^4}\eta^2, \quad \mathcal{A}_{12}(0,\eta)=-\frac{3De}{H(0)^2}\eta, \quad \mathcal{A}_{22}=1,
\label{Conformation BC OB}
\end{equation}
with $H(0)\equiv1$ under the non-dimensionalization (\ref{ND_variables_OB}).
The first two boundary conditions in (\ref{BC OB}) represent the no-slip and no-penetration conditions along the channel walls, respectively. The third condition represents the symmetry boundary condition at the centerline, and the last condition represents the integral mass
conservation along the channel. In addition, we assume a fully developed unidirectional
Poiseuille flow in the straight entry channel, so that the boundary conditions in (\ref{Conformation BC OB}) represent the corresponding conformation tensor components at $x=0$~\citep{boyko2022pressure,boyko2024flow}.

\subsection{Non-dimensional pressure drop across the non-uniform region}

In this subsection, we provide a closed-form expression for the pressure drop of an Oldroyd-B fluid across a slowly varying channel. \citet{boyko2024flow} and \citet{hinch2024fast} presented the full derivation, and therefore, we do not reproduce it here.
Specifically, they showed that, in the lubrication limit, one can calculate the pressure gradient and pressure drop without solving directly for the velocity field. First, the expression for the pressure gradient is~\cite{boyko2024flow}
\begin{equation}
\frac{dP}{dX} = -\frac{3(1 - \beta_p)}{H(X)^3} 
+ \frac{3 \beta_p}{2 De} \int_0^1 (1 - \eta^2) \left[ 
\frac{1}{H(X)} \frac{\partial ( H(X) \mathcal{A}_{11} ) }{\partial X} 
+ \frac{1}{H(X)} \frac{\partial \mathcal{A}_{12}}{\partial \eta} 
\right] d\eta.
\label{dPdX_OB_general}
\end{equation}
Integrating \eqref{dPdX_OB_general} with respect to $X$ from $0$ to $1$ and using integration by parts, we obtain the pressure drop $\Delta P= \Delta P(0)-\Delta P(1)$ across the non-uniform region (for the full derivation, see (2.17)--(2.21) in \cite{boyko2024flow})
\begin{align}
\Delta P &=  \underbrace{\, 3(1 - \beta_p) \int_{0}^{1} \frac{dX}{H(X)^3}}_{\text{\rm Solvent stress}}   +\underbrace{- \frac{3 \beta_p}{De} \int_{0}^{1} 
\frac{1}{H(X)} \left( \int_{0}^{1} \eta \mathcal{A}_{12} \, d\eta \right) dX}_{\text{\rm Elastic shear stress}}
 \nonumber \\
&+\underbrace{\frac{3 \beta_p}{2 \, De} \int_{0}^{1} (1 - \eta^2) 
\Big[ \mathcal{A}_{11} \Big]^{X=0}_{X=1} d\eta - \frac{3 \beta_p}{2 \, De} \int_{0}^{1} 
\frac{H'(X)}{H(X)} \left( \int_{0}^{1} (1 - \eta^2) \mathcal{A}_{11} \, d\eta \right) dX}_{\text{\rm Elastic normal stress}},
\label{dP_OB_general}
\end{align}
where $[ \mathcal{A}_{11}]^{X=0}_{X=1}=\mathcal{A}_{11}(0,\eta)-\mathcal{A}_{11}(1,\eta)$ and prime indicates a derivative with respect to $X$.

Equation (\ref{dP_OB_general}) resembles the result of an application of the reciprocal theorem
previously derived for the non-dimensional pressure drop of the flow of a viscoelastic fluid in a slowly
varying channel~\citep{boyko2021RT,boyko2022pressure}.
The first term on the right-hand side of
(\ref{dP_OB_general}) represents the viscous contribution of the Newtonian solvent to the pressure drop.
The second term represents
the elastic contribution due to shear stresses within the fluid domain of the non-uniform
channel. The third term represents the contribution of the elastic normal stress difference at the
inlet and outlet of the non-uniform channel. Finally, the last term represents the contribution of
the elastic normal stresses that arise due to the spatial variations in the channel shape; notably, this term vanishes for a straight channel.

\section{Lubrication analysis in the ultra-dilute limit: \texorpdfstring{$\beta_p\ll1$}{}}\label{Ultra-dilute section}

To facilitate theoretical analysis, in addition to using the orthogonal curvilinear coordinates defined in \eqref{curvilinear mapping}, we consider the ultra-dilute limit, $\beta_p \ll 1$~\citep{remmelgas1999computational,moore2012weak,li2019orientation,mokhtari2022birefringent,boyko2024flow,hinch2024fast,mahapatra2025viscoelastic}. The ultra-dilute limit represents a one-way coupling between the velocity and pressure and the elastic stresses (conformation tensor). At the leading order in $\beta_p$, the velocity of the viscoelastic fluid is parabolic, similar to a Newtonian fluid, and is unaffected by elastic stresses. The elastic stresses (i.e., spatial evolution of the conformation tensor) arise from this parabolic velocity profile. This one-way coupling enables us to derive closed-form asymptotic expressions for the conformation tensor at $O(\beta_p^{0})$ and the pressure drop for arbitrary values of the Deborah number up to $O(\beta_p)$~\citep{boyko2024flow,hinch2024fast,mahapatra2025viscoelastic}. 
To this end, we seek solutions of the form
\begin{equation}
\begin{pmatrix}
U \\
V \\
P \\
\mathcal{A}_{11} \\
\mathcal{A}_{12} \\
\mathcal{A}_{22}
\end{pmatrix}
=
\begin{pmatrix}
U_{0} \\
V_{0} \\
P_0 \\
\mathcal{A}_{11,0} \\
\mathcal{A}_{12,0} \\
\mathcal{A}_{22,0}
\end{pmatrix}
+ \beta_p \,
\begin{pmatrix}
U_{1} \\
V_{1} \\
P_1 \\
\mathcal{A}_{11,1} \\
\mathcal{A}_{12,1} \\
\mathcal{A}_{22,1}
\end{pmatrix}
+ O(\epsilon^2,\beta_p^2),
\label{asymptotic series beta}
\end{equation}
and in the following subsections, we provide closed-form asymptotic expressions for 
the leading-order conformation tensor components
and the dimensionless pressure drop up to $O(\beta_p)$.

\subsection{Velocity and conformation tensor components at the leading order in \texorpdfstring{$\beta_p$}{}}

The leading-order behavior in $\beta_p$ corresponds to a Newtonian fluid flow, so that the velocity field and pressure drop are~\citep{boyko2024flow}
\begin{equation}
U_0 =\frac{F(\eta)}{H(X)} =\frac{3}{2}\frac{1}{H(X)}(1 - \eta^2), \quad V_0 \equiv 0, \quad \text{and} \quad \Delta P_0 = 3 \int_{0}^{1} \frac{dX}{H(X)^3},
\label{U0 V0 dP0 OB}
\end{equation}
where we define $F(\eta)$ as $F(\eta) = (3/2)(1 - \eta^2)$. In orthogonal curvilinear coordinates, the velocity in the $\eta$-direction, $V_0$, 
is identically zero at the leading order in $\beta_p$. However, in the Cartesian coordinates, the vertical velocity $U_{y}$ has a non-vanishing expression $U_{y}=(3/2)H'(X)Y(H(X)^{2}-Y^{2})/H(X)^{4}$. The absence of velocity in the $\eta$-direction simplifies the theoretical analysis and enables us to derive closed-form expressions for the components of the conformation tensor. 

In the constriction region, under the ultra-dilute limit, it is convenient to use the following rescaling for the conformation tensor components~\citep{hinch2024fast,boyko2024flow}
\begin{equation}
b_{11}(X,\eta)= \frac{\mathcal{A}_{11,0}}{18De^2\eta^2/H(X)^2}, \quad b_{12}(X,\eta)= -\frac{\mathcal{A}_{12,0}}{3\eta De} , \quad b_{22}(X,\eta)=\frac{\mathcal{A}_{22,0}}{H(X)^2}.
\label{bij_highDe_ultra_OB}
\end{equation}
Substituting \eqref{asymptotic series beta} and \eqref{bij_highDe_ultra_OB} into (\ref{Conformation OB}c--e), and considering the leading order in $\beta_p$, the equations for the conformation tensor components of the Oldroyd-B fluid simplify to
\begin{subequations} 
\begin{align}
F(\eta) \frac{\partial b_{11}}{\partial X} &= -\frac{H(X)}{De} (b_{11} - b_{12}) \quad \text{with }\quad b_{11}(0,\eta) = 1, \\
F(\eta) \frac{\partial b_{12}}{\partial X} &= -\frac{H(X)}{De} (b_{12} - b_{22}) \quad \text{with }\quad b_{12}(0,\eta) = 1, \\
F(\eta) \frac{\partial b_{22}}{\partial X} &= -\frac{H(X)}{De} \left( b_{22} - \frac{1}{H(X)^2} \right) \quad \text{with }\quad b_{22}(0,\eta) = 1.
\end{align}
\label{bij_ultradilute_OB}\end{subequations}
We note that, in the ultra-dilute limit, the fully coupled system of differential equations (\ref{Conformation OB}c--e) simplifies to a set of one-way coupled partial differential equations~\eqref{bij_ultradilute_OB}.
This one-way coupling allows us to solve first for $b_{22}$, then for $b_{12}$, and finally for $b_{11}$. Solving \eqref{bij_ultradilute_OB}, we obtain closed-form expressions for $b_{22}$, $b_{12}$, and $b_{11}$ for arbitrary values of $De$ and and the shape function $H(X)$
\begin{subequations}
\begin{align}
    b_{22}&=\text{exp}\left(\frac{\mathcal{G}(X,F(\eta))}{De}\right)\left[1+ \int_0^X \text{exp}\left(-\frac{\mathcal{G}(X',F(\eta))}{De}\right)\, \frac{1}{DeF(\eta)H(X')}dX'\right],
    \label{b22_Gen}
\\
    b_{12}&=\text{exp}\left(\frac{\mathcal{G}(X,F(\eta))}{De}\right)\left[1+ \int_0^X \text{exp}\left(-\frac{\mathcal{G}(X',F(\eta))}{De}\right)\, \frac{b_{22}H(X')}{DeF(\eta)}dX'\right],
    \label{b12_Gen}
\\
    b_{11}&=\text{exp}\left(\frac{\mathcal{G}(X,F(\eta))}{De}\right)\left[1+ \int_0^X \text{exp}\left(-\frac{\mathcal{G}(X',F(\eta))}{De}\right)\, \frac{b_{12}H(X')}{DeF(\eta)}dX'\right],
    \label{b11_Gen}
\end{align}
\label{bij_Gen}\end{subequations}
where $\mathcal{G}(X,F(\eta))$ is defined as
\begin{equation}
    \mathcal{G}(X,F(\eta))=-\int_0^X \frac{H(X')}{F(\eta)} \, dX'=-\int_0^X \frac{2H(X')}{3(1-\eta^2)} \, dX'.
    \label{G}
\end{equation}
We note that our expressions for $b_{11}, b_{12}$, and $b_{22}$ are in agreement with the expressions (3.8)--(3.10) previously
derived by~\citet{boyko2024flow}. Using the rescaling \eqref{bij_highDe_ultra_OB} and the fact that $U_0=\partial U_0/\partial X=0$ at the wall $\eta=\pm 1$, the rescaled conformation tensor components at the walls of the non-uniform channel take the following form
\begin{equation}
    b_{22}^{\rm wall}=b_{12}^{\rm wall}= b_{11}^{\rm wall}=\frac{1}{H(X)^2}.
\end{equation}
In Appendix~\ref{AppB}, we derive closed-form expressions for the conformation tensor components in the uniform exit
channel for arbitrary values of the Deborah number in the ultra-dilute limit. Furthermore, we provide the asymptotic expressions for the conformation tensor components in the constriction and exit channel in the low- and high-$De$ limits.

\subsection{Pressure drop and elastic stress contributions to the pressure drop at the first order in~\texorpdfstring{$\beta_p$}{}}
\label{Pressure drop OB}

In the ultra-dilute limit, it is possible to calculate the pressure drop at $O(\beta_p)$ using the leading-order velocity field and elastic stresses. Substituting \eqref{asymptotic series beta} into \eqref{dP_OB_general}, we obtain the pressure drop at $O(\beta_p)$~\cite{boyko2024flow,hinch2024fast}
\begin{align}
\Delta P_1 = &-3\int_{0}^{1} \frac{dX}{H(X)^3} 
+ \frac{3}{2 \, De} \int_{0}^{1} (1 - \eta^2) 
\Big[\mathcal{A}_{11,0}\Big]_{X=1}^{X=0}
 d\eta  
 - \frac{3 }{2 \, De} \int_{0}^{1} 
\frac{H'(X)}{H(X)} \left( \int_{0}^{1} (1 - \eta^2) \mathcal{A}_{11,0} \, d\eta \right) dX \nonumber \\
& - \frac{3}{De} \int_{0}^{1} 
\frac{1}{H(X)} \left( \int_{0}^{1} \eta \mathcal{A}_{12,0} \, d\eta \right) dX,
\label{dP1_OB_general}
\end{align}
where the elastic normal and shear stresses $\mathcal{A}_{11,0}$ and $\mathcal{A}_{12,0}$ are obtained from (\ref{b11_Gen}) and (\ref{b12_Gen}) using~\eqref{bij_highDe_ultra_OB}.

Similar to \eqref{dP_OB_general}, the pressure drop in \eqref{dP1_OB_general} can be decomposed into its three main contributions.
The elastic normal stress (NS) contribution to the pressure drop at $O(\beta_p)$ is
\begin{equation}
\Delta P_1^{\rm NS}=\frac{3}{2 \, De} \int_{0}^{1} (1 - \eta^2) 
\Big[\mathcal{A}_{11,0}\Big]_{X=1}^{X=0}
 d\eta  - \frac{3}{2 \, De} \int_{0}^{1} 
\frac{H'(X)}{H(X)} \left( \int_{0}^{1} (1 - \eta^2) \mathcal{A}_{11,0} \, d\eta \right) dX, 
\label{dP1_NS_OB_general}
\end{equation}
and the elastic shear stress (SS) contribution to the pressure drop at $O(\beta_p)$ is
\begin{equation}
    \Delta P_1^{\rm SS}=- \frac{3}{De} \int_{0}^{1} 
\frac{1}{H(X)} \left( \int_{0}^{1} \eta \mathcal{A}_{12,0} \, d\eta \right) dX.
\label{dP1_SS_OB_general}
\end{equation}
Thus, the dimensionless
pressure drop $\Delta P= \Delta p /(\mu_{0}q\ell/2h_{0}^{3}) $ as a function of the shape function $H(X)$, the Deborah number $De$, and the viscosity ratio $\beta_p\ll 1$, up to $O(\beta_p)$, is given by
\begin{gather}
    \Delta P= \Delta P_0 + \beta_p \Delta P_1 +O(\epsilon^2,\beta_p^2),
    \label{dP total OB low-beta}
\end{gather}
where the expressions for $\Delta {P_{0}}$ and $\Delta {P_{1}}$ are given in (\ref{U0 V0 dP0 OB}) and (\ref{dP1_OB_general}), respectively. In Appendix~\ref{AppC}, we provide the low-$De$ asymptotic expressions for the pressure drop of an Oldroyd-B fluid in the constriction. In the following subsection, we present the results for the pressure drop at high Deborah numbers in the ultra-dilute limit.

\subsubsection{Pressure drop at \texorpdfstring{$O(\beta_p)$}{} in the high-\texorpdfstring{$De$}{} limit}

We here provide expressions for the pressure drop and the elastic normal and shear stress contributions to the
pressure drop at high Deborah numbers. 
Substituting (\ref{bij_ultradilute_OB_highDe_sol}) into (\ref{dP1_NS_OB_general}) and (\ref{dP1_SS_OB_general}), and using (\ref{bij_highDe_ultra_OB}), the elastic normal and shear stress contributions to the pressure drop at $O(\beta_p)$ in the high-$De$ limit are
\refstepcounter{equation}
$$
\Delta P_1^{\rm NS} = \frac{9 De}{5}\left(\frac{1}{H(0)^{2}} - \frac{1}{H(1)^{2}} \right)
 \quad \text{and} \quad 
\Delta P_1^{ \rm SS} = 3 \int_{0}^{1} \frac{dX}{H(X)}.\eqno{(\theequation{\text{a,b}})}\label{dP_highDe_ultra_simplified}
$$
Therefore, using \eqref{dP1_OB_general}, \eqref{dP total OB low-beta}, and \eqref{dP_highDe_ultra_simplified}, the total pressure drop across the non-uniform channel ($H(0) \neq H(1)$) in the high-$De$ limit is
\begin{equation}
    \Delta P= \underbrace{\, 3(1 - \beta_p) \int_{0}^{1} \frac{dX}{H(X)^3}}_{\text{\rm Solvent stress}} + \underbrace{3\beta_p\int_{0}^{1} \frac{dX}{H(X)}}_{\text{\rm Elastic shear stress}}
    +\underbrace{\frac{9}{5}\beta_p De\left(\frac{1}{H(0)^{2}} - \frac{1}{H(1)^{2}} \right)}_{\text{\rm Elastic normal stress}} \; \text{for} \; De \gg 1.
\label{dP_nonuniform_HighDe}
\end{equation}
For the symmetric constriction, we have $H(0)=H(1)=1$, and thus the elastic stress contributions to the
pressure drop, \eqref{dP_highDe_ultra_simplified}, simplify to
\refstepcounter{equation}
$$
\Delta P_1^{\rm NS} = 0
 \quad \text{and} \quad 
\Delta P_1^{ \rm SS} = 3 \int_{0}^{1} \frac{dX}{H(X)}.\eqno{(\theequation{\text{a,b}})}\label{dP_highDe_constr}
$$
From \eqref{dP_highDe_constr}, it follows that, in the high-$De$ limit, the elastic normal stress contribution to the pressure drop vanishes in the constriction channel. Finally, substituting \eqref{dP_highDe_constr} into \eqref{dP1_OB_general} and \eqref{dP total OB low-beta} provides the total pressure drop across the \emph{constriction} in the high-$De$ limit
\begin{equation}
    \Delta P= \underbrace{\, 3(1 - \beta_p) \int_{0}^{1} \frac{dX}{H(X)^3}}_{\text{\rm Solvent stress}} + \underbrace{3\beta_p\int_{0}^{1} \frac{dX}{H(X)}}_{\text{\rm Elastic shear stress}}
    \; \text{for} \; De \gg 1.
    \label{dP_nonuniform_HighDe constriction}
\end{equation}
Similar to the contraction geometry, the elastic shear stresses in the constriction are lower than their fully relaxed value, $\mathcal{A}_{12} = -3De\eta$, due to insufficient distance to approach this fully relaxed value in the constriction, as we show in the inset of Fig.~\ref{F3}(c). As a result, the elastic shear stress contribution to the pressure drop, $3 \beta_p \int_0^1 H(X)^{-1}dX$, is smaller than the steady Poiseuille value, $3 \beta_p \int_0^1 H(X)^{-3}dX$, thus reducing the total pressure drop~\cite{boyko2024flow,hinch2024fast}. 
However, unlike the contraction case ($H(0)>H(1)$), where elastic normal stresses lead to a reduction in the pressure drop~\cite{boyko2024flow,hinch2024fast,mahapatra2025viscoelastic}, the contribution from elastic normal stresses vanishes in the constriction geometry, where $H(0) = H(1)$.

\section{Results and discussion}\label{Results}

\begin{figure}
    \centering
    \includegraphics[scale=1.2]{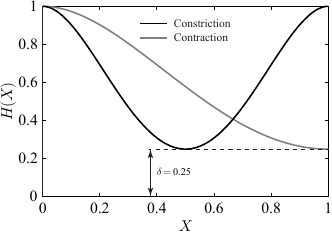}
    \caption{Illustration of the smooth constriction and contraction geometries considered in this work. The black curve represents the constriction shape function~\eqref{H_constriction} and the gray curve represents the contraction shape function~\eqref{H_contraction}. All calculations were performed using $\delta=0.25$.}
    \label{F2}
\end{figure}

In this section, we present the theoretical results for the elastic stresses and pressure drop of the Oldroyd-B fluid flowing through a constriction in the ultra-dilute limit. In this work, we present the results for $\beta_p=0.05$.
As an illustrative example, we consider the case of a smooth quartic polynomial contraction-expansion (quartic constriction) of the form
\begin{equation}
    H(X) = 1 -  2^4(1-\delta)X^2 (1 - X)^2 \quad 0 \leq X \leq 1.
\label{H_constriction}
\end{equation}
In (\ref{H_constriction}), the constriction ratio $\delta$ ($0<\delta\leq1$) represents the minimum half-height of the channel at $X = 0.5$, such that $H(0.5) = \delta$. 
For the quartic constriction, the
entry and exit half-heights are $H(X=0)=H(X=1)=1$.
This constriction shape function is symmetric about $X=0.5$ with vanishing slope at the inlet and outlet, $H'(X=0)=H'(X=1)=0$, as represented in Fig.~\ref{F2} by the black curve. In this work, we present the results for $\delta=0.25$.

To highlight the similarities and differences between constriction and contraction channels, we compare and contrast the predictions for the pressure drop across the constriction (\ref{H_constriction}) with that across a smooth contraction of the form
\begin{equation}
    H(X) = 1 -  (1-\delta) X^2 (2 - X)^2 \quad 0 \leq X \leq 1.
\label{H_contraction}
\end{equation}
The contraction shape function (\ref{H_contraction}) has vanishing slope at the inlet and outlet, $H'(X=0)=H'(X=1)=0$, and satisfies $H(X=0)=1$ and $H(X=1)=\delta$, as represented in Fig.~\ref{F2} by the gray curve.  We note that both shape functions, (\ref{H_constriction}) and (\ref{H_contraction}), yield identical Newtonian pressure drop, $3 \int_0^1 H(X)^{-3}dX=48.225$, for $\delta=0.25$. Furthermore, both shapes yield $\Delta P_1^{ \rm SS}=3 \int_0^1 H(X)^{-1}dX=6.248$, and thus, as given in (\ref{dP_highDe_constr}b), for these shapes, we obtain the same elastic shear stress contribution to the pressure drop at high Deborah numbers.

\subsection{Streamwise variation of elastic stresses in the constriction and exit channel}

\begin{figure}
    \centering
    \includegraphics[scale=1.2]{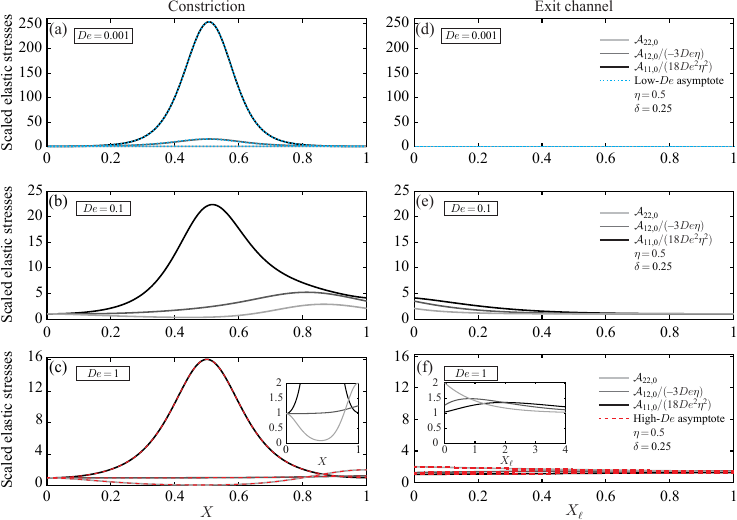}
    \caption{The streamwise variation of leading-order elastic stresses on $\eta=0.5$ in a slowly varying constriction and
exit channel in the ultra-dilute limit. (a--c) Scaled elastic stresses $\mathcal{A}_{11,0}/(18De^2\eta^2)$, $\mathcal{A}_{12,0}/(-3De\eta)$ and $\mathcal{A}_{22,0}$ in the constriction as a function of $X$ for (a) $De=0.001$, (b) $De=0.1$, and (c) $De=1$. (d--f) Scaled elastic
stresses in the exit channel $\mathcal{A}_{11,0}/(18De^2\eta^2)$, $\mathcal{A}_{12,0}/(-3De\eta)$ and $\mathcal{A}_{22,0}$ in the exit channel as a function of $X_\ell$ for (d) $De=0.001$, (e) $De=0.1$, and (f) $De=1$. The insets in (c) and (f) show the zoom-in view. Solid lines represent the semi-analytical solutions \eqref{b22_Gen}--\eqref{b11_Gen} (constriction) and (\ref{A_general_exit}) (exit channel). Cyan dotted lines represent the low-$De$ asymptotic solutions (\ref{A_ij_lowDe}) (constriction)
and (\ref{A_general_exit}) with (\ref{Aij_ref_LowDe_OB_Ultra_pol4}a--c) (exit channel). 
Red dashed lines represent the high-$De$ asymptotic solutions (\ref{bij_ultradilute_OB_highDe_sol}) (constriction) and (\ref{A_general_exit}) with (\ref{Aij_ref_HighDe_OB_Ultra}) (exit channel). All calculations were performed using $\delta=0.25$.}
    \label{F3}
\end{figure}

First, we study the spatial variation of the elastic stresses. Figure~\ref{F3} presents the streamwise variation of the leading-order elastic stresses, scaled by their entry values, on $\eta = 0.5$ in the constriction and exit channel in the ultra-dilute limit for (a,d) $De = 0.001$, (b,e) $De = 0.1$, and (c,f) $De = 1$. 
For a small Deborah number of $De = 0.001$, the elastic stresses are approximately symmetric about the minimum gap ($X = 0.5$) and return to their entry values by the end of the constriction (Fig.~\ref{F3}(a)). Consequently, we observe very small variation in the relaxation along the exit channel (Fig.~\ref{F3}(d)). Such low-$De$ behavior is similar to that observed in the contraction geometry, where the elastic stresses reach their downstream
 fully relaxed values by the end of the contraction~\cite{boyko2024flow}. 
 Furthermore, for $\delta=0.25$, the elastic shear and axial normal stresses increase by a factor
of $1/\delta^2=16$ and $1/\delta^4=256$, respectively, by the center of the constriction ($X=0.5$), while the transverse normal stress approximately retains its entry value, consistent with the low-$De$ asymptotic solutions (\ref{A_ij_lowDe}), represented by cyan dotted lines. 

When $De = 0.1$, it is evident that the elastic stresses in the constriction region exhibit a non-symmetric distribution~(Fig.~\ref{F3}(b)), in contrast to the nearly symmetric profiles observed at $De=0.001$, indicating that the elastic stresses do not fully relax by the end of the constriction. Therefore, we observe a spatial relaxation in the exit channel (Fig.~\ref{F3}(e)). 
Our findings are in qualitative agreement with recent results of~\citet{hinch2024fast}, who presented in their Fig. 14 the streamwise variation of the elastic stresses in the constriction (from $X=0$ to $X=2$ with $\delta=0.5$) and exit channel for $De=0.5$ and $c=1$, which corresponds to $\beta_p =0.5 \nll 1$.

For $De=1$, the elastic stresses become approximately symmetric about $X = 0.5$, closely resembling the behavior observed at the low-$De$ case of $De=0.001$. However, a closer look shows, as illustrated in the insets of Figs.~\ref{F3}(c) and~\ref{F3}(f), that the elastic shear and transverse normal stresses do not have enough residence time to fully relax and return to their entry values by the end of the constriction. Even though the elastic axial normal stress almost returns to its entry values by the end of the constriction, it exhibits a spatial relaxation in the exit channel.
Interestingly, in contrast to the monotonic relaxation of elastic stresses in the exit channel for $De=0.1$, we observe that for $De=1$, the elastic shear and axial normal stresses exhibit a non-monotonic relaxation behavior.

Furthermore, there is excellent agreement between the semi-analytical results (solid lines) and the high-$De$ asymptotic solutions (\ref{bij_ultradilute_OB_highDe_sol}) and (\ref{A_general_exit}) with (\ref{Aij_ref_HighDe_OB_Ultra}) (red dashed lines). Such excellent agreement for $De=1$ in the constriction (\ref{H_constriction}) is consistent with the findings of \citet{boyko2024flow} for the contraction channel (\ref{H_contraction}). Nevertheless,  when comparing the results shown in Fig.~\ref{F3} for the constriction with the corresponding results for the contraction (see Fig. 3 in \cite{boyko2024flow}), we observe that the elastic stresses in the exit channel of the constriction relax over a significantly shorter distance compared to those in the contraction, consistent with the discussion by~\citet{szabo1997start}.
In subsection~\ref{Pressure gradient relaxation}, we analyze the necessary exit length for constriction and contraction channels by examining the downstream relaxation of the pressure gradient and quantifying the distance required for it to fully relax.

\subsection{Pressure drop in the constriction at low and high Deborah numbers}

\begin{figure}
    \centering
    \includegraphics[scale=1.2]{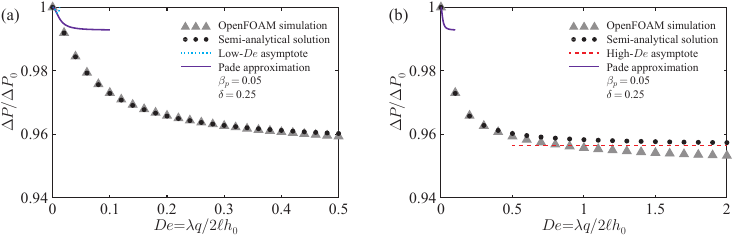}
    \caption{Non-dimensional pressure drop at low and high Deborah numbers for the Oldroyd-B fluid in a constriction channel in the ultra-dilute limit. (a,b) Scaled pressure drop $\Delta P/ \Delta P_0$ as a function of $De=\lambda q/(2\ell h_0)$. Gray triangles represent the OpenFOAM simulation results. Black dots represent the semi-analytical solution \eqref{dP total OB low-beta}. The cyan dotted line represents the low-$De$ asymptotic solution \eqref{dP total OB low-De}.
The purple solid line represents the low-$De$ Padé approximation \eqref{Pade-approximation OB low-De}. The red dashed line represents the high-$De$ asymptotic solution~\eqref{dP_nonuniform_HighDe constriction}. All calculations were performed using $\beta_p=0.05$ and $\delta=0.25$.}
    \label{F4}
\end{figure}

In this subsection, we study the pressure drop behavior of the Oldroyd-B fluid across the constriction at low and high Deborah numbers. We first consider the pressure drop at relatively low Deborah numbers and compare the results of the low-$\beta_p$ lubrication analysis, applicable for all $De$, with low-$De$ asymptotic expressions for pressure drop derived by~\citet{boyko2022pressure},~\citet{housiadas2023lubrication}, and~\citet{mahapatra2025viscoelastic}.
For completeness, in Appendix~\ref{AppC}, we provide the low-$De$ asymptotic expressions for the pressure drop of the Oldroyd-B fluid up to $O(De^4)$, along with the corresponding low-$De$ Padé approximation.

We present in Fig.~\ref{F4}(a) the scaled pressure drop $\Delta P/ \Delta P_0$ as a function of $De=\lambda q/(2\ell h_0)$ for the Oldroyd-B fluid in a constriction channel \eqref{H_constriction} in the ultra-dilute
limit. Black dots represent the
semi-analytical solution \eqref{dP total OB low-beta}, the cyan dotted line represents the low-$De$ asymptotic solution \eqref{dP total OB low-De}, and the purple solid line represents the low-$De$ Padé approximation \eqref{Pade-approximation OB low-De}.
In addition, we validate the predictions of our semi-analytical and low-$De$ asymptotic results against the two-dimensional finite-volume numerical simulations, represented by gray triangles. 
The details of the numerical implementation in the finite-volume software OpenFOAM are provided in Appendix~\ref{AppD}.

First, it is evident from Fig.~\ref{F4}(a) that, at low Deborah numbers, the dimensionless pressure drop monotonically decreases with increasing $De$, consistent with previous numerical studies using the Oldroyd-B model to investigate the viscoelastic fluid flow in constriction geometries~\cite{szabo1997start,binding2006contraction,aguayo2008excess,hinch2024fast}. Second, while there is excellent agreement between the predictions of the low-$\beta_p$ lubrication analysis with $\beta_p=0.05$ and the OpenFOAM simulations, we observe that, in the constriction, the low-$De$ asymptotic solution and even the low-$De$ Padé approximation cannot accurately capture the pressure drop except for very low values of $De$. This weak agreement is somewhat surprising, given the excellent accuracy of the low-$De$ Padé approximation in capturing the pressure drop for a 4:1 hyperbolic contraction, as shown in recent studies; see, for example, Fig. 4(b) in~\cite{mahapatra2025viscoelastic} and Fig. 4(c) in~\cite{sialmas2025general}. One possible reason for this weak agreement is the absence of odd powers of $De$ in the low-$De$ asymptotic expression for the pressure drop in the smooth and symmetric constriction, as discussed in Appendix~\ref{AppC}, which limits the accuracy of the Padé approximation in capturing the correct pressure drop behavior.

Next, in Fig.~\ref{F4}(b), we present the scaled pressure drop $\Delta P/ \Delta P_0$ as a function of $De=\lambda q/(2\ell h_0)$ at high Deborah numbers up to $De=2$ in the ultra-dilute limit. We observe excellent
agreement between the predictions of the low-$\beta_p$  lubrication analysis with $\beta_p=0.05$ (black dots)  and the
OpenFOAM simulations (gray triangles), resulting in a relative error of less than $0.5\%$ up to $De=2$. In contrast to a monotonic pressure drop reduction with $De$ observed at low Deborah numbers, the pressure drop of the Oldroyd-B fluid in the constriction levels off to a plateau at high Deborah numbers, as accurately captured by the high-$De$ asymptotic solution~\eqref{dP_nonuniform_HighDe constriction}, represented by the red dashed line. In the following subsections, we elucidate the origin of the high-$De$ plateau and further contrast our findings with results obtained for the contraction.

We note that~\citet{hinch2024fast} studied the pressure drop of an Oldroyd-B fluid in a slowly varying constriction similar to~\eqref{H_constriction}. They solved numerically the lubrication equations (\ref{Conformation OB}) subject to the boundary conditions (\ref{BC OB})--(\ref{Conformation BC OB}) for $c=1$. However, these authors did not observe the leveling off of the pressure drop with increasing $De$ in their Fig. 13.
We rationalize this behavior by noting that~\citet{hinch2024fast} considered $c = 1$, corresponding to $\beta_p = 0.5$, which does not represent the ultra-dilute limit. Furthermore, their results for the pressure drop were limited to $De\lesssim0.35$ for $\delta = 0.25$.
In fact, as noted by~\citet{hinch2024fast}, since their constriction extended from $X = 0$ to $X = 2$, the effective Deborah number should be halved, so that $De \lesssim 0.175$ for $\delta = 0.25$, which clearly does not correspond to high Deborah numbers. Therefore, not surprisingly, the high-$De$ asymptotic solution~\eqref{dP_nonuniform_HighDe constriction}, which is based on the ultra-dilute limit, could not capture the numerically calculated pressure drop in this case.

\begin{figure}
    \centering
    \includegraphics[scale=1.2]{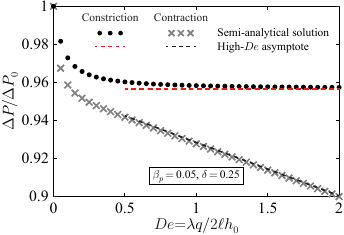}
    \caption{Comparison of non-dimensional pressure drop for the Oldroyd-B fluid in constriction and contraction channels in the ultra-dilute limit. Scaled pressure drop $\Delta P/ \Delta P_0$ as a function of $De=\lambda q/(2\ell h_0)$. Black dots and gray crosses represent the semi-analytical solutions~\eqref{dP total OB low-beta} for the constriction and contraction geometries. 
    The red dashed line represents the high-$De$ asymptotic solution (\ref{dP_nonuniform_HighDe constriction}) for the constriction.
    The black dashed line represents the high-$De$ asymptotic solution (\ref{dP_nonuniform_HighDe}) for the contraction. All calculations were performed using $\beta_p=0.05$ and $\delta=0.25$.}
    \label{F5}
\end{figure}

\subsection{Pressure drop and elastic stress contributions to the pressure drop: constriction versus contraction}

As both constriction and contraction geometries are commonly encountered in different fluidic systems and applications, it is of particular interest to compare and contrast their pressure drop behavior. We present in Fig.~\ref{F5} the scaled pressure drop $\Delta P/ \Delta P_0$ of the Oldroyd-B fluid in the constriction  (\ref{H_constriction}) and contraction (\ref{H_contraction}) as a function of $De=\lambda q/(2\ell h_0)$ for $\beta_p=0.05$.
Black dots and gray crosses represent the semi-analytical solutions~\eqref{dP total OB low-beta} for the constriction and contraction geometries. 
The red dashed line represents the high-$De$ asymptotic solution (\ref{dP_nonuniform_HighDe constriction}) for the constriction.
The black dashed line represents the high-$De$ asymptotic solution (\ref{dP_nonuniform_HighDe}) for the contraction. 

At low Deborah numbers, the pressure drop of the Oldroyd-B fluid monotonically decreases with $De$ in both constriction and contraction geometries. However, at high Deborah numbers, unlike a plateau value in the constriction, for the contraction, the pressure drop of the Oldroyd-B fluid monotonically decreases with $De$ and scales linearly
with $De$. Such a linear pressure drop reduction in the contraction is consistent with the previous studies considering the ultra-dilute limit~\citep{boyko2024flow,hinch2024fast,mahapatra2025viscoelastic} and is accurately captured by the high-$De$ asymptotic solution, represented by the black dashed line.
Furthermore, in agreement with prior studies on planar contractions, we observe that the semi-analytical solution for the pressure drop in both constriction and contraction geometries approaches the high-$De$ asymptotic solutions for order-one Deborah numbers.

\begin{figure}[t]
    \centering
    \includegraphics[scale=1.2]{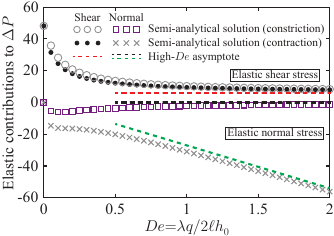}
    \caption{Elastic contributions to the non-dimensional pressure drop of the Oldroyd-B fluid, scaled by $\beta_p$, as a function of~$De=\lambda q/(2\ell h_0)$ in the constriction and contraction channels in the ultra-dilute limit. Gray circles and black dots represent the semi-analytical solutions (\ref{dP1_SS_OB_general}) for elastic shear stress contribution in the constriction and contraction geometries. Purple squares and gray crosses represent the semi-analytical solutions (\ref{dP1_NS_OB_general}) for elastic normal stress contribution in the constriction and contraction geometries. The red dashed line represents the high-$De$ asymptotic solution~(\ref{dP_highDe_ultra_simplified}b) for elastic shear stress contribution in the constriction and contraction geometries. Black and green dashed lines represent the high-$De$ asymptotic solutions (\ref{dP_highDe_constr}a) and (\ref{dP_highDe_ultra_simplified}a) for elastic normal stress contribution in the constriction and contraction geometries. All calculations were performed using $\delta=0.25$.}
\label{F6}
\end{figure}

To understand the origin of the difference in pressure drop behavior between constriction and contraction channels, we examine the relative importance of elastic contributions to the pressure drop in both geometries. 
We present in Fig.~\ref{F6} the elastic contributions to the non-dimensional pressure drop of the Oldroyd-B fluid across the constriction and contraction channels, scaled by $\beta_p$, as a function of~$De=\lambda q/(2\ell h_0)$ in the ultra-dilute limit. 
Gray circles and black dots represent the elastic shear stress contribution in the constriction and contraction geometries obtained from the semi-analytical solution (\ref{dP1_SS_OB_general}).
Purple squares and gray crosses represent the elastic normal stress contribution in the constriction and contraction geometries obtained from the semi-analytical solution (\ref{dP1_NS_OB_general}).
The red dashed line represents the elastic shear stress contribution in the constriction and contraction geometries obtained from the high-$De$ asymptotic solution~(\ref{dP_highDe_ultra_simplified}b).
Black and green dashed lines represent the elastic normal stress contribution in the constriction and contraction geometries obtained from the high-$De$ asymptotic solutions (\ref{dP_highDe_constr}a) and (\ref{dP_highDe_ultra_simplified}a), respectively. 
Clearly, there is excellent agreement between our high-$De$ asymptotic solutions and the semi-analytical predictions, consistent with the previous results shown in Fig.~\ref{F5}. 

The first source for the pressure drop reduction is the elastic shear stress contribution. 
For both the constriction and contraction channels, the elastic shear stress contribution monotonically decreases with increasing $De$, eventually reaching a plateau at high Deborah numbers, which is accurately captured by the high-$De$ asymptotic solution~(\ref{dP_highDe_ultra_simplified}b). We note that for both shape functions, (\ref{H_constriction}) and (\ref{H_contraction}), we obtain the same elastic shear stress contribution to the pressure drop at high Deborah numbers, namely $\Delta P_1^{ \rm SS}=3 \int_0^1 H(X)^{-1}dX=6.248$. The observed reduction in the elastic shear stress contribution to the pressure drop arises from the elastic shear stresses in the constriction and contraction being lower than their fully relaxed value, $\mathcal{A}_{12} = -3De\eta/H(1)^2$, due to insufficient time (or distance) to approach their eventual relaxed values by the end of the constriction/contraction. As a result, the elastic shear stress contribution to the pressure drop at high Deborah numbers, given by $3 \beta_p \int_0^1 H(X)^{-1}dX$, is smaller than the steady Poiseuille value, $3 \beta_p \int_0^1 H(X)^{-3}dX$, thereby reducing pressure drop~\cite{boyko2024flow,hinch2024fast}.

The second source for the pressure drop reduction is the elastic normal stress
contribution. For the contraction, the elastic normal stress
contribution monotonically decreases with increasing $De$ and scales linearly with $De$ at high Deborah numbers. As noted by~\citet{hinch2024fast} and~\citet{boyko2024flow}, the observed decrease in pressure drop is due to the increased elastic normal stresses at the end of the contraction compared to the beginning. These higher elastic normal stresses pull the fluid forward, thereby reducing the pressure needed to drive the flow.

In sharp contrast to the contraction, the elastic normal stress contribution in the constriction first slightly decreases at low Deborah numbers and then levels off to zero at high Deborah numbers, as accurately captured by the high-$De$ asymptotic solution~(\ref{dP_highDe_constr}a). This behavior arises because, at high Deborah numbers in the ultra-dilute limit, the elastic normal stresses are symmetric about the minimum gap, as shown in Fig.~\ref{F3}(c), thereby eliminating their net contribution to the pressure drop. 
Therefore, while at low and moderate Deborah numbers two distinct physical mechanisms contribute to the pressure drop reduction in both the constriction and contraction channels, at high $De$, a key difference emerges: in the contraction channel, both elastic normal and shear stresses contribute to the pressure drop reduction, whereas in the constriction channel, the reduction is solely due to the elastic shear stresses.

\subsection{Pressure gradient relaxation in the exit channel: constriction versus contraction}\label{Pressure gradient relaxation}

In this subsection, we study the relaxation of the pressure gradient in the exit channel following the constriction and contraction, and compare the distances required for the spatial relaxation of the pressure gradient. Substituting $H(X)=H(1)$ into~\eqref{dPdX_OB_general}, we obtain the pressure gradient in the exit channel
\begin{equation}
\frac{dP}{dX_\ell} = -\frac{3(1 - \beta_p)}{H(1)^3} + \frac{3 \beta_p}{2De} 
\int_{0}^{1} (1 - \eta^2) \frac{\partial \mathcal{A}_{11}}{\partial X} d\eta 
+ \frac{3 \beta_p}{H(1) De} \int_{0}^{1} \eta \mathcal{A}_{12} d\eta.
\label{dP/dX_Exit_OB}
\end{equation}
The latter expression (\ref{dP/dX_Exit_OB}) can be expressed as~\cite{boyko2024flow},
\begin{equation}
\left( \frac{dP}{dX_\ell} + \frac{3}{H(1)^3} \right) \frac{1}{\beta_p} = 
\frac{3}{H(1)^3} - \frac{H(1)}{De^2} \int_{0}^{1} \mathcal{A}_{11} d\eta 
- \frac{3}{H(1) De} \int_{0}^{1} \eta \mathcal{A}_{12} d\eta,
\label{Scaled dP/dX_Exit_OB}
\end{equation}
where $(dP/dX_{\ell}+ 3/{H(1)^3})/\beta_p$ is the scaled pressure gradient. We note that, in the ultra-dilute limit, the right-hand side of (\ref{Scaled dP/dX_Exit_OB}) is independent of $\beta_p$, and the elastic normal and shear stresses, $\mathcal{A}_{11}$ and $\mathcal{A}_{12}$, in the exit channel are calculated using the expressions given in Appendix~\ref{AppB}.

\begin{figure}
    \centering
    \includegraphics[scale=1.2]{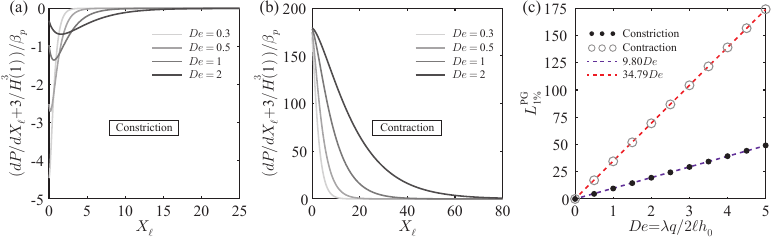}
    \caption{The spatial relaxation of the pressure gradient for the Oldroyd-B fluid in the uniform exit channel of constriction and contraction geometries in the ultra-dilute limit. (a) Scaled pressure gradient $(dP/dX_{\ell}+ 3/{H(1)^3})/\beta_p$ in the exit channel of a constriction as a function of the downstream distance $X_\ell$ for $De=0.3,0.5,1$, and $2$. 
    (b) Scaled pressure gradient $(dP/dX_{\ell}+ 3/{H(1)^3})/\beta_p$ in the exit channel of a contraction as a function of the downstream distance $X_\ell$ for $De=0.3,0.5,1$, and $2$.  
    Solid lines represent the semi-analytical solutions obtained from \eqref{Scaled dP/dX_Exit_OB} using \eqref{A_general_exit}.~(c) Non-dimensional relaxation length $L_{1\,\%}^{\mathrm{PG}}$, defined as the downstream distance where the magnitude of the scaled pressure gradient falls below $1\%$ of its maximum value, as a function of $De$ in the exit channel following the constriction (black dots) and the contraction (gray circles). The purple dashed line is $9.8De$ and the red dashed line is $34.79De$. All calculations were performed using $\delta=0.25$.}
    \label{F7}
\end{figure}

We present in Figs.~\ref{F7}(a,b) the scaled pressure gradient in the exit channel of constriction
and contraction geometries, as a function of the downstream distance $X_\ell$ for $De=0.3,0.5,1$, and $2$. 
In the exit channel of the contraction, the scaled pressure gradient relaxes exponentially over the downstream distance. However, in the exit channel of the constriction, we observe a non-monotonic but still exponential relaxation for $De \ge 0.5$.
This non-monotonic relaxation of the pressure gradient arises from the non-monotonic relaxation of the elastic shear and axial normal stresses, as illustrated in Fig.~\ref{F3}(f). 

For both exit channels following constriction and contraction, the scaled pressure gradient relaxes over a longer distance as $De$ increases, consistent with previous studies~\cite{debbaut1988numerical,alves2003benchmark,boyko2024flow,hinch2024fast}.
To quantify the length of the exit channel required for the scaled pressure gradient relaxation, we define the relaxation length, $L_{1\,\%}^{\mathrm{PG}}$, as the downstream distance over which the magnitude of the scaled pressure gradient decays to $1\%$ of its maximum value. 
We present in Fig.~\ref{F7}(c) the relaxation length $L_{1\,\%}^{\mathrm{PG}}$
as a function of $De$ in the exit channel following the constriction (black dots) and
the contraction (gray circles).
It is evident from Fig.~\ref{F7}(c) that in both exit channels following constriction and contraction, the relaxation length increases linearly with $De$ throughout the
investigated range of Deborah numbers. More specifically, it follows from Fig.~\ref{F7}(c)
that the relaxation length in the exit channel of the contraction is approximately $L_{1\,\%}^{\mathrm{PG}}\approx (34.79 \pm 0.04)\times De$ (red dashed line), in close agreement with the expression, $L_{1\%}^{\mathrm{PG}}\approx (5.3 \pm 0.5)\times (3De/2H(1))$, reported by~\citet{boyko2024flow}. Indeed, for the contraction we have $H(1)=\delta=0.25$, so that the latter expression yields 
$L_{1\%}^{\mathrm{PG}}\approx (31.8 \pm 3)\times De$. On the other hand, we obtain that the relaxation length in the exit channel of the constriction is approximately $L_{1\,\%}^{\mathrm{PG}}\approx (9.80 \pm 0.03)\times De$ (purple dashed line). Clearly, the relaxation length in the exit channel of the constriction is much shorter than in the exit channel of the contraction, consistent with the discussion by~\citet{szabo1997start}.

\section{Concluding remarks}\label{Concluding_remarks}

\begin{table}[t]
  \begin{center}
\def~{\hphantom{0}}
\renewcommand{\arraystretch}{1.1}
  \begin{tabular}{lcc}
 & Constriction & Contraction  \\
\hline
\hline
Pressure drop at low $De$ & Monotonic pressure drop reduction & Monotonic pressure drop reduction \\
Pressure drop at high $De$ & Leveling off to a  plateau & Linear decrease with $De$ \\

Elastic normal stress contribution 
at high $De$ & Approaching zero & Linear decrease with $De$ \\

Elastic shear stress contribution 
at high $De$ & Leveling off to a  plateau & Leveling off to a  plateau \\

Elastic stress relaxation in the exit channel  & Monotonic relaxation at low $De$ & Monotonic relaxation  \\
& Non-monotonic relaxation at high $De$ & \\

Relaxation length in the exit channel, $L_{1\,\%}^{\mathrm{PG}}$ & $\approx (9.80 \pm 0.03)\times De$ (shorter) & $\approx (34.79 \pm 0.04)\times De$ (longer) \\
\end{tabular}
\caption{Comparative summary of theoretical predictions for the pressure-driven flow of an Oldroyd-B fluid through slowly varying constriction and contraction channels in the ultra-dilute limit.}\label{Table_1}
\end{center}
\end{table}

In this work, we have studied the pressure-driven flow of an Oldroyd-B fluid in a slowly varying constriction channel at low and high Deborah numbers. 
Employing lubrication theory in orthogonal curvilinear coordinates and considering the ultra-dilute limit, we used the one-way coupling between the parabolic velocity and elastic stresses to derive closed-form expressions for the polymer conformation tensor and pressure drop for arbitrary values of the Deborah number. We further provided the corresponding asymptotic expressions for the elastic stresses and the pressure drop in the low- and high-Deborah-number limits for the constriction
and exit channels.
We validated our semi-analytical and asymptotic results for the non-dimensional pressure drop in the smooth constriction with two-dimensional finite-volume numerical simulations for the Oldroyd-B fluid and found excellent agreement. In addition, we studied the relaxation of the elastic stresses and pressure gradient in the exit channel following the constriction. Finally, we compared and contrasted our theoretical predictions for the constriction channel with the previous results of~\citet{boyko2024flow} for the contraction channel, highlighting their differences and similarities. In particular, we showed that the pressure gradient in the exit channel of the constriction relaxes over a much shorter distance compared to that in the contraction. As a result, the contraction channel necessitates a longer exit channel than the constriction channel.

Table~\ref{Table_1} presents a comparison of theoretical predictions for the pressure-driven flow of an Oldroyd-B fluid through slowly
varying constriction and contraction channels in the ultra-dilute limit, summarizing our key findings. 
For both constriction and contraction channels, the pressure drop of the Oldroyd-B fluid monotonically decreases with $De$ at low Deborah numbers. However, at high Deborah numbers, in contrast to a linear decrease with $De$ for the contraction, the pressure
drop across the constriction levels off to a plateau (Fig.~\ref{F5}). We identified the physical mechanisms that are responsible for
such pressure drop behavior in both geometries (Fig.~\ref{F6}). At low and moderate Deborah numbers, both elastic normal and shear stresses contribute to the pressure drop reduction in the constriction and contraction channels. However, at high Deborah numbers, a significant difference arises. In the contraction channel, both elastic normal and shear stresses continue to contribute to the pressure drop reduction. In contrast, in the constriction channel, the reduction is exclusively due to elastic shear stresses, as the contribution from elastic normal stresses vanishes. 

In this study, we described the viscoelastic behavior of the fluid using the Oldroyd-B model, which is the simplest form of the elastic dumbbell model combining both viscous and elastic stresses. 
While the Oldroyd-B model provides valuable insights into the hydrodynamics within the constriction and contraction channels, it has several shortcomings~\citep{beris2021continuum,hinch2021oldroyd,shaqfeh2021oldroyd,sanchez2022understanding,StoneShelleyBoyko2023}. Therefore, as a future research direction, it would be interesting to study more complex elastic dumbbell models. Such models could incorporate additional microscopic features of polymer solutions, including the finite extensibility of polymer chains, the conformation-dependent
friction coefficient and the conformation-dependent non-affine deformation~\citep{phan1984study,BoykoStone2024Perspective,boyko2025pressureFD}.

To date, the theoretical analysis of viscoelastic channel flows has primarily focused on simple geometries such as straight channels, contractions, expansions, and constrictions, using lubrication theory. A natural and interesting extension of these studies is to analyze viscoelastic flows in complex channel networks composed of multiple slowly varying segments of diverse geometries, including straight and non-uniform sections. Such configurations are more representative of realistic microfluidic systems and biological environments. 
However, even in simple geometries, finite-volume and finite-element simulations encounter accuracy and convergence issues above a certain non-large Deborah number due to the high-Weissenberg-number problem~\citep{owens2002computational,alves2021numerical}. Therefore, we expect that studying viscoelastic flows in complex channel networks through numerical simulations will be challenging, if not impossible, beyond low Deborah numbers.
In contrast, we believe that even for more complex geometries, our theoretical framework, which relies on lubrication theory and the ultra-dilute limit, will enable the development of a simplified model, allowing us to study the viscoelastic flow in such channel networks at non-small Deborah numbers, highlighting its fundamental and practical importance.

\begin{acknowledgments}
We gratefully acknowledge support from the Israel Science Foundation (Grant No. 1942/23). B.M.\ acknowledges partial support from the Department of Science and Technology (DST), India, through the INSPIRE Faculty Fellowship (Grant No.\ DST/INSPIRE/04/2024/003069). E.B.\ acknowledges support from the Israeli Council for Higher Education Yigal Alon Fellowship. 

\end{acknowledgments}

\section*{Author Contributions}

E.B. conceived, designed, and supervised the project and acquired the funding that supported this research. Y.K. and E.B. performed the theoretical research. B.M. performed the finite-volume numerical simulations. E.B., Y.K., and B.M. analyzed the results and wrote the manuscript. 

\appendix

\section{Non-dimensional lubrication equations in Cartesian coordinates}\label{AppA}

Using the non-dimensionalization (\ref{ND_variables_OB})$-$(\ref{De and Wi}),
to the leading order in $\epsilon$, the governing equations (\ref{Continuity+Momentum})--(\ref{Evolution OB})
take the form
\begin{subequations} \label{ND Cart}
\begin{gather}
\frac{\partial U_{x}}{\partial X}+\frac{\partial U_{y}}{\partial Y}=0, \label{Continuity ND Cart} \\
\frac{\partial P}{\partial X}=(1-\beta_p)\frac{\partial^{2}U_{x}}{\partial Y^{2}}+\frac{\beta_p}{De}\left(\frac{\partial\mathcal{A}_{xx}}{\partial X}+\frac{\partial\mathcal{A}_{xy}}{\partial Y}\right), \label{Momentum z ND Cart} \\
\frac{\partial P}{\partial Y}=0, \label{Momentum y ND Cart} \\
U_{x}\frac{\partial \mathcal{A}_{xx}}{\partial X}+U_{y}\frac{\partial \mathcal{A}_{xx}}{\partial Y}-2\frac{\partial U_{x}}{\partial X}\mathcal{A}_{xx}-2\frac{\partial U_{x}}{\partial Y}\mathcal{A}_{xy}=-\frac{1}{De}\mathcal{A}_{xx} ,\label{Axx ND Cart} \\
U_{x}\frac{\partial \mathcal{A}_{xy}}{\partial X}+U_{y}\frac{\partial \mathcal{A}_{xy}}{\partial Y}-\frac{\partial U_{y}}{\partial X}\mathcal{A}_{xx}-\frac{\partial U_{x}}{\partial Y}\mathcal{A}_{yy}=-\frac{1}{De}\mathcal{A}_{xy} ,\label{Axy ND Cart} \\
U_{x}\frac{\partial \mathcal{A}_{yy}}{\partial X}+U_{y}\frac{\partial \mathcal{A}_{yy}}{\partial Y}-2\frac{\partial U_{y}}{\partial X}\mathcal{A}_{xy}-2\frac{\partial U_{y}}{\partial Y}\mathcal{A}_{yy}=-\frac{1}{De}(\mathcal{A}_{yy}-1). \label{Ayy ND Cart}
\end{gather}\end{subequations}
From (\ref{Momentum y ND Cart}), it follows that
$P=P(X)$, i.e., the pressure is independent of $Y$ up to $O(\epsilon^2)$, consistent with the classical lubrication approximation. We note that the scaled $\mathcal{A}_{xx}$ on the right-hand side of (\ref{Axx ND Cart}) relaxes to $\epsilon^2$, which is neglected at the leading order in $\epsilon$.

\section{Elastic stresses in the exit channel and asymptotic expressions in the low- and high-\texorpdfstring{$De$}{} limits}\label{AppB}

In this Appendix, we calculate the elastic stresses in the exit channel and present the asymptotic expressions for the conformation tensor components in the constriction and exit channel in the low- and high-$De$ limits.
In the exit channel, the velocity field and the pressure drop at the leading order in $\beta_p$ are given by~\cite{boyko2024flow}
\begin{equation}
U_0 =  \frac{3}{2}\frac{1}{H(1)}(1 - \eta^2), \quad V_0 \equiv 0, \quad \text{and} \quad \Delta P_0 = \frac{3 L}{H(1)^3} .
\label{U0 V0 dP0 exit OB}
\end{equation}
Using \eqref{U0 V0 dP0 exit OB}, and considering the leading order in $\beta_p$, the equations for the conformation tensor in the straight exit channel ($H\equiv H(1)$) for the Oldroyd-B fluid reduce to 
\begin{subequations}
    \begin{gather}
        U_0 \frac{\partial \mathcal{A}_{22,0}}{\partial X} 
        = -\frac{1}{De} (\mathcal{A}_{22,0} - 1),
    \\
        U_0 \frac{\partial \mathcal{A}_{12,0}}{\partial X}
        -\frac{1}{H(1)} \frac{d U_0}{d \eta} \mathcal{A}_{22,0}
        = -\frac{1}{De} \mathcal{A}_{12,0}, 
   \\
        U_0 \frac{\partial \mathcal{A}_{11,0}}{\partial X}
        - \frac{2}{H(1)} \frac{d U_0}{d \eta} \mathcal{A}_{12,0}
        = -\frac{1}{De} \mathcal{A}_{11,0}.
    \end{gather}\label{Conformation exit ultra OB}\end{subequations}
Similar to \eqref{bij_ultradilute_OB}, \eqref{Conformation exit ultra OB} are one-way coupled equations, allowing us to solve first for $\mathcal{A}_{22,0}$, then for $\mathcal{A}_{12,0}$, and finally for $\mathcal{A}_{11,0}$.
Solving \eqref{Conformation exit ultra OB}, we obtain closed-form expressions for the spatial relaxation of the conformation tensor components in the exit channel~\cite{boyko2024flow} 
\begin{subequations}
\begin{align}
\mathcal{A}_{22,0} &= 1 + (\mathcal{A}^{\text{ref}}_{22,0}(\eta) - 1) \exp\left(-\frac{2 H(1)X_\ell}{3De(1 - \eta^2)}\right), \label{eq:B3} \\
\mathcal{A}_{12,0} &= -\frac{3De}{H(1)^2}\eta + \exp\left(-\frac{ 2 H(1) X_\ell}{3De(1 - \eta^2)}\right) 
\left[\mathcal{A}^{\text{ref}}_{12,0}(\eta) + \frac{3De}{H(1)^2}\eta - \frac{2\eta (\mathcal{A}^{\text{ref}}_{22,0}(\eta) - 1)X_\ell}{H(1)(1 - \eta^2)}\right], \label{eq:B4} \\
\mathcal{A}_{11,0} &= \frac{18De^2}{H(1)^4} \eta^2 + \exp\left(-\frac{ 2 H(1) X_\ell}{3De(1 - \eta^2)}\right) 
\Bigg[\mathcal{A}^{\text{ref}}_{11,0}(\eta) -\frac{18De^2}{H(1)^4} \eta^2 \notag \\
 &\quad + \frac{4\eta^2 (\mathcal{A}^{\text{ref}}_{22,0}(\eta) - 1)X_\ell^2}{ H(1)^2(1 - \eta^2)^2}
- \frac{4\eta X_\ell [3De\eta + H(1)^2 \mathcal{A}^{\text{ref}}_{12,0}(\eta)]}{H(1)^3(1 - \eta^2)} \Bigg].
\label{eq:B5}
\end{align}
\label{A_general_exit}\end{subequations}
In \eqref{A_general_exit}, we have introduced the rescaled axial coordinate $X_\ell = X - 1$ and the reference distributions of the conformation tensor components at the outlet of the constriction ($X = 1$), $\mathcal{A}^{\text{ref}}_{22,0}(\eta) = \mathcal{A}_{22,0}(X = 1, \eta)$, $\mathcal{A}^{\text{ref}}_{12,0}(\eta) = \mathcal{A}_{12,0}(X = 1, \eta)$, and $\mathcal{A}^{\text{ref}}_{11,0}(\eta) = \mathcal{A}_{11,0}(X = 1, \eta)$.

\subsection{Elastic stresses in the constriction and exit channel: low~\texorpdfstring{$De$}{}}

At low Deborah numbers, we solve equations~\eqref{bij_ultradilute_OB} iteratively for the conformation tensor components in the constriction region to obtain~\cite{boyko2024flow}
\begin{subequations}
\begin{align}
 \mathcal{A}_{22,0} &= 1 + \frac{3 De H'}{H^2}(1 - \eta^2) 
+ \frac{9 De^2 [4H'^2 - H H'']}{2 H^4}(1 - \eta^2)^2 + \frac{27 De^3 [24 H'^3 - 13 H H' H'' + H^2 H''']}{4 H^6}(1 - \eta^2)^3,
\label{A22_lowDe} \\
 \mathcal{A}_{12,0} &= - \frac{3 De}{H^2} \eta 
- \frac{18 De^2 H'}{H^4} \eta (1 - \eta^2)
- \frac{81 De^3 [4H'^2 - H H'']}{2 H^6} \eta (1 - \eta^2)^2,
\label{A12_lowDe} \\
 \mathcal{A}_{11,0} &= \frac{18 De^2}{H^4} \eta^2
+ \frac{162 De^3 H'}{H^6} \eta^2 (1 - \eta^2)
+ \frac{486 De^4 [4H'^2 - H H'']}{H^8} \eta^2 (1 - \eta^2)^2.
\label{A11_lowDe}
\end{align}
\label{A_ij_lowDe}\end{subequations}
We note that expressions~\eqref{A_ij_lowDe} are consistent with the results of~\citet{boyko2024flow}. In addition, we have extended the low-Deborah-number lubrication analysis of~\citet{boyko2024flow} to higher asymptotic orders and obtained the conformation tensor components up to $O(De^5)$. However, the resulting expressions are lengthy and, thus, not presented here.

To calculate the elastic stresses in the exit channel, we first calculate the reference distributions of the conformation tensor components at the beginning of the exit channel. Using \eqref{A_ij_lowDe}, we obtain $\mathcal{A}^{\text{ref}}_{22}(\eta)$, $\mathcal{A}^{\text{ref}}_{12}(\eta)$, and $\mathcal{A}^{\text{ref}}_{11}(\eta)$ in the low-$De$ limit
\begin{subequations}
\begin{align}
\mathcal{A}^{\text{ref}}_{22,0}=&1 - \frac{9De^2 H''(1)}{2H(1)^3 }(1 - \eta^2)^2 + \frac{27De^3 H'''(1)}{4 H(1)^4} (1 - \eta^2)^3  \notag \\  & + \frac{81De^4[13H''(1)^2-H(1)H^{(4)}(1)] }{8H(1)^6} (1 - \eta^2)^4-   \frac{10935De^5 H''(1)H'''(1)}{16H(1)^7}(1 - \eta^2)^5,
\\ 
\mathcal{A}^{\text{ref}}_{12,0}= &-\frac{3De }{H(1)^2}\eta
+ \frac{81De^3 H''(1)}{2H(1)^5}\eta (1 - \eta^2)^2 + \frac{81De^4 H'''(1)}{H(1)^6}\eta (1 - \eta^2)^3  \notag \\  &+\frac{729 De^5[13H''(1)^2- H(1)H^{(4)}(1)]}{8H(1)^8}\eta (1 - \eta^2)^4  ,\\ 
\mathcal{A}^{\text{ref}}_{11,0}= & \frac{18De^2}{H(1)^4}\eta^2 -\frac{486 De^4 H''(1)}{H(1)^7}\eta^2(1-\eta^2)^2 +\frac{243De^5 H'''(1)}{H(1)^8}\eta^2(1-\eta^2)^3 , 
\end{align}
\label{Aij_ref_LowDe_OB_Ultra_pol4}\end{subequations}
where, for a smooth geometry, we have assumed that $H'(1)=H^{(5)}(1)=0$. We note that the expressions for the reference distributions of the conformation tensor components in \eqref{Aij_ref_LowDe_OB_Ultra_pol4} are consistent with those presented by \citet{boyko2024flow}, and further extend the calculation to include terms up to $O(De^5)$.
Substituting \eqref{Aij_ref_LowDe_OB_Ultra_pol4} into \eqref{A_general_exit} provides the conformation tensor components at low Deborah numbers in the exit channel.  The resulting expressions are lengthy and, thus, not presented
here.

\subsection{Elastic stresses in the constriction and exit channel: high~\texorpdfstring{$De$}{}}

At high Deborah numbers, we solve equations~\eqref{bij_ultradilute_OB} iteratively for the conformation tensor components in the central core of the constriction region where $F(\eta) \gg 1/De$ to obtain
\begin{subequations}\begin{align}
b_{22} &= 1 + \frac{1}{De F(\eta)} I_1 - \frac{1}{(De F(\eta))^2} I_2 + \frac{1}{(De F(\eta))^3} I_3 + O((De F(\eta))^{-4}), \\
b_{12} &= 1 + \frac{1}{(De F(\eta))^2} I_2 - \frac{2}{(De F(\eta))^3} I_3 + O((De F(\eta))^{-4}), \\
b_{11} &= 1 + \frac{1}{(De F(\eta))^3} I_3 + O((De F(\eta))^{-4}),
\end{align}
\label{bij_ultradilute_OB_highDe_sol}\end{subequations}
where
\begin{subequations} 
\begin{align}
I_1 &= \int_0^X \frac{1 - H^2(X')}{H(X')} \, dX', \\
I_2 &= \int_0^X H(X') \int_0^{X'} \frac{1 - H^2(X'')}{H(X'')} \, dX'' \, dX', \\
I_3 &= \int_0^X H(X') \int_0^{X'} H(X'') \int_0^{X''} \frac{1 - H^2(X''')}{H(X''')} \, dX''' \, dX'' \, dX'.
\end{align}
\label{I1_I2_I3}\end{subequations}
Equations \eqref{bij_ultradilute_OB_highDe_sol} and \eqref{I1_I2_I3} have been previously derived by \citet{hinch2024fast}. Unfortunately, there are misprints in the prefactor of the last term of (\ref{bij_ultradilute_OB_highDe_sol}b) in their corresponding expression, which should be $2$ and not $1$.

Similar to the low-$De$ limit, at high Deborah numbers, using the rescaling (\ref{bij_highDe_ultra_OB}), we calculate the conformation tensor components at the beginning of the exit channel to obtain the elastic stresses in the exit channel
\begin{subequations}
\begin{align}
 \mathcal{A}^{\text{ref}}_{22,0}&= H(1)^2\left(1+\frac{1}{DeF(\eta)}I_1(1) -\frac{1}{(DeF(\eta))^{2}}I_2(1)+ \frac{1}{(DeF(\eta))^{3}}I_3(1)\right),\\
 \mathcal{A}^{\text{ref}}_{12,0}&= -3\eta De \left( 1 +\frac{1}{(DeF(\eta))^{2}}I_2(1)- \frac{2}{(DeF(\eta))^{3}}I_3(1)\right)
 ,\\ 
 \mathcal{A}^{\text{ref}}_{11,0}&= \frac{18  \eta^2De^2}{H(1)^2}\left( 1+ \frac{1}{(DeF(\eta))^{3}}I_3(1)\right),
\end{align}
\label{Aij_ref_HighDe_OB_Ultra}\end{subequations}
where $I_1,I_2,I_3$ are given in \eqref{I1_I2_I3}, respectively.
Substituting the reference distributions of the conformation tensor components \eqref{Aij_ref_HighDe_OB_Ultra} into \eqref{A_general_exit}, we obtain the conformation tensor components in the exit channel for $De \gg1$.
We note that the asymptotic expressions for the elastic stresses 
for $De\gg 1$ are valid within the core flow region. Similar to the case of the contraction, these
expressions do not hold near the walls of the channel, where a viscoelastic boundary layer of $O(De^{-1})$ thickness exists~\citep{boyko2024flow, hinch2024fast}. 
Nevertheless, as pointed out by~\citet{hinch2024fast} and~\citet{boyko2024flow}, this boundary
layer contributes only a small $O(\beta_p De^{-1})$ correction to the pressure drop. Indeed, as shown in Figs.~\ref{F4}(b) and~\ref{F5}, the excellent agreement between the semi-analytical results
and the high-$De$ asymptotic solution (\ref{dP_nonuniform_HighDe constriction}), based on the components of the conformation tensor within the core flow region, indicates that the boundary layer near the walls makes a negligible contribution to the pressure drop in the constriction.

\section{Asymptotic expressions for the pressure drop  in the constriction in the low-\texorpdfstring{$De$}{} limit}\label{AppC}

In this appendix, we provide asymptotic expressions for the pressure drop of the Oldroyd-B fluid in the low-Deborah-number limit developed by~\citet{boyko2022pressure},~\citet{housiadas2023lubrication}, and~\citet{mahapatra2025viscoelastic}. To calculate the pressure drop, these authors expanded the velocity, pressure
drop, and conformation tensor components into perturbation series in powers of the Deborah number
$De \ll1$ and solved order by order the resulting equations. 
\citet{boyko2022pressure} provided explicit expressions for the non-dimensional pressure drop of the Oldroyd-B fluid in the low-$De$ limit up to $O(De^3)$.~\citet{housiadas2023lubrication} 
extended the low-Deborah-number lubrication analysis of \citet{boyko2022pressure} up to $O(De^8)$ and provided explicit expressions for the pressure drop up to $O(De^4)$. \citet{mahapatra2025viscoelastic}~recently derived similar expressions for the dimensionless pressure drop of the Oldroyd-B and FENE-CR fluids up to $O(De^4)$. 

The contraction geometry studied by~\citet{boyko2022pressure} and \citet{housiadas2023lubrication} satisfies $H(0) > H(1)$, leading to the pressure drop contributions that include both odd and even powers of $De$. In contrast, the symmetry of the constriction channel \eqref{H_constriction} causes the pressure drop corrections to appear only at even powers of $De$.
Specifically, the first-order correction in $De$, $\Delta P^{(1)}$, vanishes because $H(0) = H(1)$, and the third-order correction, $\Delta P^{(3)}$, vanishes as $H(0) = H(1)$ and $H''(0) = H''(1)$ (see ~(4.8) and~(4.14) in~\cite{boyko2022pressure}).

As expected, the pressure drop at the leading order in $De$ is the Newtonian pressure drop 
\begin{gather}
     \Delta P^{(0)}=3\int_{0}^{1}\frac{dX}{H(X)^3}.
     \label{P_0_ND_OB}
\end{gather}
The second-order and fourth-order corrections to the pressure drop for the Oldroyd-B fluid are given by~\cite{boyko2022pressure,housiadas2023lubrication,mahapatra2025viscoelastic}
\begin{gather}
\Delta P^{(2)} = \frac{324}{35} \, \beta_p \int_0^1 \left( \frac{14 H'(X)^2}{H(X)^7} - \frac{3 H''(X)}{H(X)^6} \right) \, dX,
\label{dP2 OB}
\end{gather}
and
\begin{align}
\Delta P^{(4)}=&\int_0^1 \Bigg(
a_1\frac{\, H''(X)^2}{H(X)^9}
+a_2  \frac{\, H'''(X) H'(X)}{H(X)^9}
+ a_3 \frac{\, H(X)^3 H^{''''}(X)}{H(X)^{11}}\notag \\
    &+ a_4 \frac{\,H(X)'^4}{H(X)^{11}} 
+ a_5 \frac{\, H'(X)^2 H''(X)}{H(X)^{10}}
\Bigg) dX  + a_6\Bigg(\frac{H'''(0)}{H(0)^8}-\frac{H'''(1)}{H(1)^8}\Bigg),
\label{dP4 OB}
\end{align}
where the coefficients $a_1,...,a_{6}$ for the symmetric constriction shape \eqref{H_constriction} are summarized in Table~\ref{tab:OB_coeffs}.
\begin{table}
  \begin{center}
\def~{\hphantom{0}}
\renewcommand{\arraystretch}{2.2}
  \begin{tabular}{cccc}
Coefficient & Expression & Coefficient & Expression \\
\hline
\hline
$a_1$  & $\dfrac{3240\beta_p}{13\,475} \left[\beta_p(41 - 70\beta_p) + 910\right]$ & $a_2$ & $\dfrac{4536\beta_p}{13\,475} \left[\beta_p(119 - 82\beta_p) + 750\right]$\\

$a_3$  & $\dfrac{1296\beta_p}{13\,475} \left[2\beta_p(7\beta_p - 5) - 175\right]$ & $a_4$ & $\dfrac{9072\beta_p}{13\,475} \left[11\beta_p(83 - 40\beta_p) + 2400\right]$\\

$a_5$  & $\dfrac{1944\beta_p}{13\,475} \left[\beta_p(1666\beta_p - 2789) - 12\,950\right]$ &
$a_6$  & $\dfrac{5184\beta_p}{13\,475} \left[3\beta_p(7\beta_p-24) + 175\right]$
\end{tabular}
\caption{Coefficients appearing in \eqref{dP4 OB} for the fourth-order pressure drop $\Delta P^{(4)}$ of the Oldroyd-B fluid in a planar constriction channel.}\label{tab:OB_coeffs}
\end{center}
\end{table}

The expressions for $\Delta {P^{(0)}}$, $\Delta {P^{(2)}}$, and $\Delta {P^{(4)}}$, given by (\ref{P_0_ND_OB}), 
(\ref{dP2 OB}),
and (\ref{dP4 OB}), determine the dimensionless
pressure drop $\Delta P= \Delta p /(\mu_{0}q\ell/2h_{0}^{3})$ of the Oldroyd-B fluid as a function of the shape function $H(X)$, the
viscosity ratio $\beta_p$, and the Deborah number up to $O(De^4)$,
\begin{equation}
    \Delta P= \Delta P^{(0)} + De^2 \Delta P^{(2)} + De^4 \Delta P^{(4)} +O(\epsilon^2,De^5).
    \label{dP total OB low-De}
\end{equation}

Having the low-$De$ expressions for $\Delta {P^{(0)}}$, $\Delta {P^{(2)}}$, and $\Delta {P^{(4)}}$, it is possible to accelerate the convergence of the asymptotic series \eqref{dP total OB low-De}
using the diagonal Pad\'{e} [1/1] approximation, given by~\cite{housiadas2017improved}
\begin{equation}
\Delta P_{ \rm Pade}=\Delta P^{(0)}+\frac{De^{2}\Big(\Delta P^{(2)}\Big)^2}{\Delta P^{(2)}-De^{2}\Delta P^{(4)}},
\label{Pade-approximation OB low-De}
\end{equation}
where the small parameter in \eqref{Pade-approximation OB low-De} is $De^2$, rather than $De$. We note that, unlike the diagonal Padé [2/2] approximation used by~\citet{mahapatra2025viscoelastic} for the contraction geometry, in the case of the quartic constriction, the diagonal Padé [2/2] reduces to the [1/1] form, since both $\Delta P^{(1)}$ and $\Delta P^{(3)}$ vanish due to symmetry.

We emphasize that the low-$De$ expressions for the pressure drop (\ref{P_0_ND_OB})--(\ref{dP4 OB}) are valid for $\beta_p \nll 1$. Nevertheless, we use a small value of $\beta_p = 0.05$ to allow a consistent comparison between the low-$De$ asymptotic solution and our low-$\beta_p$ semi-analytical solution.

\section{Details of OpenFOAM numerical simulations}\label{AppD}

To validate our theoretical predictions based on lubrication theory and the ultra-dilute limit, we perform two-dimensional finite-volume simulations for the viscoelastic flow of the Oldroyd-B fluid in a constriction channel. Specifically, we solve the system of
nonlinear governing equations (\ref{Continuity+Momentum})--(\ref{Evolution OB}) using an open-source framework OpenFOAM~\citep{jasak2007openfoam}  integrated with the viscoelastic flow solver \textsc{RheoTool}~\citep{pimenta2017stabilization} employing the log-conformation formulation to ensure numerical stability at high Deborah numbers~\citep{pimenta2017stabilization,habla2014numerical,kumar2021elastic,fattal2004constitutive,fattal2005time}. Further details of the numerical implementation in the finite-volume software OpenFOAM are provided in~\citet{mahapatra2025viscoelastic}.

We consider a constriction geometry with an aspect ratio $\epsilon = h_0/\ell = 0.02$,  a constriction ratio $\delta=0.25$, and a viscosity ratio $\beta_p = \mu_p/\mu_0 = 0.05$, corresponding to the ultra-dilute 
limit. In all simulations, we set both the entrance and exit lengths to match the length of the non-uniform region, resulting in $\ell_0 = \ell_{\ell} = \ell$. We use a non-dimensional time step of $\Delta T = 10^{-6}$ for both low- and high-$De$ numerical simulations, where the time is non-dimensionalized using the residence time in the constriction $t_c = \ell/u_c = 1$ s.
To assess the grid sensitivity, we have performed simulations with two different mesh resolutions, containing 51686 and 132966 node points, for Deborah numbers ranging from 0 to 2 in increments of 0.1, and established grid independence with a
maximum relative error of 0.2\% for the pressure drop.

\bibliography{literature}

\end{document}